\documentclass[aps,prb,reprint,superscriptaddress]{revtex4-2}

\usepackage[T1]{fontenc}
\usepackage{mathptmx}

\usepackage{graphicx}

\usepackage[version=4]{mhchem}
\usepackage{textcomp}
\usepackage{dcolumn}
\usepackage{bm}
\usepackage{esdiff}
\usepackage{natbib}
\usepackage[colorlinks=true,citecolor=blue,linkcolor=blue,urlcolor=blue,pdfhighlight =/O]{hyperref}
\begin{document}
\title{Ionic liquid gating induced two superconductor-insulator phase transitions in spinel oxide Li$_{1 \pm x}$Ti$_2$O$_{4-\delta}$}

\author{Zhongxu Wei}
\altaffiliation{These authors contributed equally to this work.}
\affiliation{Beijing National Laboratory for Condensed Matter Physics, Institute of Physics, Chinese Academy of Sciences, Beijing 100190, China}
\affiliation{School of Physical Sciences, University of Chinese Academy of Sciences, Beijing 100049, China}

\author{Qian Li}
\altaffiliation{These authors contributed equally to this work.}
\affiliation{Beijing National Laboratory for Condensed Matter Physics, Institute of Physics, Chinese Academy of Sciences, Beijing 100190, China}

\author{Ben-Chao Gong}
\altaffiliation{These authors contributed equally to this work.}
\affiliation{Department of Physics and Beijing Key Laboratory of Opto-electronic Functional Materials \& Micro-nano Devices, Renmin University of China, Beijing 100872, China}

\author{Xinjian Wei}
\altaffiliation{These authors contributed equally to this work.}
\affiliation{Beijing National Laboratory for Condensed Matter Physics, Institute of Physics, Chinese Academy of Sciences, Beijing 100190, China}
\affiliation{School of Physical Sciences, University of Chinese Academy of Sciences, Beijing 100049, China}
\affiliation{International Center for Quantum Materials, School of Physics, Peking University, Beijing 100871, China}
\affiliation{Beijing Academy of Quantum Information Sciences, Beijing 100193, China}

\author{Wei Hu}
\author{Zhuang Ni}
\affiliation{Beijing National Laboratory for Condensed Matter Physics, Institute of Physics, Chinese Academy of Sciences, Beijing 100190, China}
\affiliation{School of Physical Sciences, University of Chinese Academy of Sciences, Beijing 100049, China}

\author{Ge He}
\affiliation{Walther Meissner Institut, Bayerische Akademie der Wissenschaften, 85748 Garching, Germany}

\author{Mingyang Qin}
\affiliation{Beijing National Laboratory for Condensed Matter Physics, Institute of Physics, Chinese Academy of Sciences, Beijing 100190, China}
\affiliation{School of Physical Sciences, University of Chinese Academy of Sciences, Beijing 100049, China}

\author{Anna Kusmartseva}
\affiliation{Department of Physics, Loughborough University, Loughborough LE11 3TU, UK}

\author{Fedor V. Kusmartsev}
\affiliation{Department of Physics, Loughborough University, Loughborough LE11 3TU, UK}
\affiliation{College of Art and Science, Khalifa University, PO Box 127788, Abu Dhabi, UAE}

\author{Jie Yuan}
\affiliation{Beijing National Laboratory for Condensed Matter Physics, Institute of Physics, Chinese Academy of Sciences, Beijing 100190, China}
\affiliation{Songshan Lake Materials Laboratory, Dongguan, Guangdong 523808, China}

\author{Beiyi Zhu}
\author{Qihong Chen}
\affiliation{Beijing National Laboratory for Condensed Matter Physics, Institute of Physics, Chinese Academy of Sciences, Beijing 100190, China}

\author{Jian-Hao Chen}
\affiliation{International Center for Quantum Materials, School of Physics, Peking University, Beijing 100871, China}
\affiliation{Key Laboratory for the Physics and Chemistry of Nanodevices, Peking University, Beijing 100871, China}
\affiliation{Beijing Academy of Quantum Information Sciences, Beijing 100193, China}

\author{Kai Liu}
\affiliation{Department of Physics and Beijing Key Laboratory of Opto-electronic Functional Materials \& Micro-nano Devices, Renmin University of China, Beijing 100872, China}

\author{Kui Jin}\email[Corresponding author: ]{kuijin@iphy.ac.cn}
\affiliation{Beijing National Laboratory for Condensed Matter Physics, Institute of Physics, Chinese Academy of Sciences, Beijing 100190, China}
\affiliation{School of Physical Sciences, University of Chinese Academy of Sciences, Beijing 100049, China}
\affiliation{Songshan Lake Materials Laboratory, Dongguan, Guangdong 523808, China}

\begin{abstract}
The associations between emergent physical phenomena (e.g., superconductivity) and orbital, charge, and spin degrees of freedom of $3d$ electrons are intriguing in transition metal compounds. Here, we successfully manipulate the superconductivity of spinel oxide Li$_{1\pm x}$Ti$_2$O$_{4-\delta}$ (LTO) by ionic liquid gating. A dome-shaped superconducting phase diagram is established, where two insulating phases are disclosed both in heavily electron-doping and hole-doping regions. The superconductor-insulator transition (SIT) in the hole-doping region can be attributed to the loss of Ti valence electrons. In the electron-doping region, LTO exhibits an unexpected SIT instead of a metallic behavior despite an increase in carrier density. Furthermore, a thermal hysteresis is observed in the normal state resistance curve, suggesting a first-order phase transition. We speculate that the SIT and the thermal hysteresis stem from the enhanced $3d$ electron correlations and the formation of orbital ordering by comparing the transport and structural results of LTO with the other spinel oxide superconductor MgTi$_2$O$_4$ (MTO), as well as analysing the electronic structure by first-principles calculations. Further comprehension of the detailed interplay between superconductivity and orbital ordering would contribute to the revealing of unconventional superconducting pairing mechanism.

\end{abstract}

\maketitle

Clarifying the competition and cooperation between superconductivity and other collective orders is a promising way to unveil the mechanism of high-temperature superconductivity \cite{Paglione2010,Jin2011,Stewart2011,Keimer2015}. Similar to the cuprate and iron-based superconductors, the superconductivity of spinel oxide Li$_{1 \pm x}$Ti$_2$O$_{4-\delta}$ (LTO) originates from $3d$ orbital electrons. Besides, the $d$-$d$ electron correlations in LTO are also remarkable \cite{Tunstall1994,Chen2011}. In contrast to Fe and Cu in superconductors, the $3d$ orbitals of Ti in LTO are much less occupied, given that the Ti-sublattice in stoichiometric LiTi$_2$O$_4$ consists of equal amounts of Ti$^{3+}$ and Ti$^{4+}$ \cite{Johnston1976}. Although the measurements of inelastic neutron scattering \cite{Green1997}, specific heat \cite{Sun2004}, and Andreev reflection spectroscopy \cite{Tang2006} provide evidences for Bardeen-Cooper-Schrieffer like superconductivity in LTO, the photoelectron spectroscopy \cite{Edwards1984} and the theoretical studies made by Anderson \emph{et al.} \cite{Anderson1987,*Anderson1987_2} prefer unconventional pairing mechanisms, such as polaron and resonating valence bond pictures. Complicated interactions among charge, orbital, and spin degrees of freedom in LTO also induce many novel phenomena, e.g., orbital-related state \cite{Jin2015,Chen2017}, pseudogap \cite{Okada2017}, anisotropic electron-phonon coupling \cite{He2017}, and anomalous upper critical field \cite{PhysRevB.100.184509}. Therefore, it is essential to elaborately depict the detailed interplay between superconductivity and related collective orders in LTO.

Carrier density is crucial for superconductivity \cite{Ye1193}, which can be effectively tuned by chemical doping. In order to dope LTO with holes, previous works have attempted to reduce the Li$^{+}$ content by exposing LTO to air \cite{Rygula1993,Moshopoulou1999} and soaking LTO in hydrochloric acid \cite{Moshopoulou1993,Moshopoulou1994}, leading to a first increase and then decrease in the zero resistance transition temperature ($T_\mathrm{c0}$). However, when holes are injected by Li$^{+}$ deintercalation and O$^{2-}$ intercalation based on ionic liquid gating (ILG), $T_\mathrm{c0}$ decreases monotonically \cite{Maruyama2015}. On the other hand, by tuning the oxygen pressure during sample fabrication, it is found that an increase in O$^{2-}$ content will induce a transition from LiTi$_2$O$_{4-\delta}$ to Li$_4$Ti$_5$O$_{12}$ along with a superconductor-insulator transition (SIT) \cite{Jia2018}. Therefore, the roles of Li$^{+}$ and O$^{2-}$ in hole-doping discussed by various approaches are distinct and desire further investigation. In contrast, the studies of electron-doped LTO are rather fewer. Typically, when electrons are introduced into LTO by ILG, the superconductivity is significantly suppressed and followed by an insulator behavior \cite{Yoshimatsu2015}. Up to now, the origin of this SIT is still an open question. Overall, it is worthy to systematically dope LTO and perform detailed measurements to clarify the issues above.

\begin{figure}[ht!]
\centering
\includegraphics[width=\linewidth]{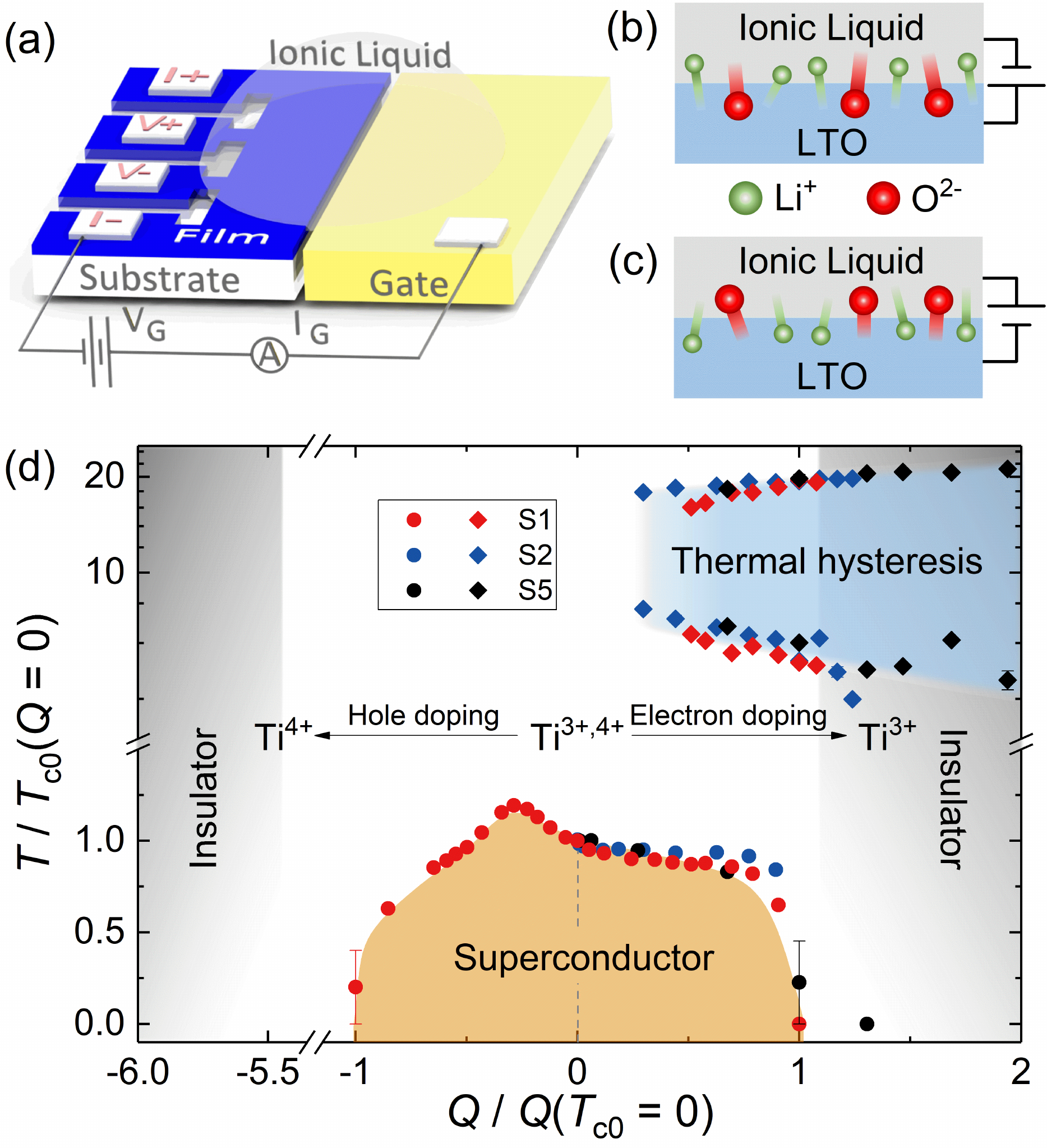}
\caption{(a) The schematic illustration of the ILG device. (b-c) The schematic illustration of ions injection and removal under a negative (b) and positive (c) gate voltage. (d) The phase diagram of LTO. The solid circles represent the normalized transition temperatures $T_\mathrm{c0}(Q)$/$T_\mathrm{c0}(Q = 0)$ for sample S1, S2, and S5. The diamonds stand for the normalized characteristic temperatures at which the warming and cooling resistance curves deviate. The transport results of sample S5 are shown in Supplemental Material \cite{supplemental}.}
\label{fig1}
\end{figure}

In this Letter, we systematically dope LTO films with electrons and holes by ILG, and successfully manipulate the superconductivity. A phase diagram of LTO is established in which a dome-shaped superconducting phase locates between two insulating phases, as shown in Fig. \ref{fig1}(d). With increasing hole-doping level, $T_\mathrm{c0}$ of LTO first increases and then drops, followed by an irreversible SIT which can be explained by the loss of Ti valence electrons. Furthermore, \emph{in situ} x-ray diffraction (XRD) measurements show that the $c$-axis lattice constant in the insulating region decreases by $5\%$, indicating that a new chemical phase may be induced. With enhancing electron-doping level, the superconductivity is gradually suppressed, and finally an unexpected reversible SIT occurs despite an increase in electron density. In addition, a thermal hysteresis appears in the temperature-dependent resistance [$R(T)$] curve of the electron-doped LTO. Different to the case of hole-doping, the lattice constant changes slightly during the whole electron-doping process. Based on the transport and structural results as well as electronic structure by first-principles calculations, we suggest that the SIT in electron-doped LTO, along with the thermal hysteresis, can be attributed to the enhanced $3d$ electron correlations and the formation of orbital ordering.


The $(00l)$-oriented LTO thin films S1-S5 are epitaxially grown on $(00l)$-oriented MgAl$_2$O$_4$ (MAO) substrates under high vacuum by pulsed laser deposition using a KrF excimer laser \cite{Jin2015,Jia2018}. The background vacuum is better than $5\times 10^{-7}$ Torr, and a sintered Li$_4$Ti$_5$O$_{12}$ ceramic target is used to fabricate the films. During deposition, the grown temperature, the laser energy desity, and repetition rate are fixed at 700\textcelsius, $\sim 2$ J/cm$^2$, and 4 Hz, respectively. The thickness of LTO films we used is $\sim 200$ nm, confirmed by scanning electron microscopy. Figure \ref{fig1}(a) shows the configuration of the ILG device, where LTO is patterned to carry out resistance measurements. In the ILG experiment, LTO and Pt electrode are covered by the ionic liquid, N,N-diethyl-N-methyl-N-(2-methoxyethyl)ammonium bis(trifluoromethylsulphonyl)imide (DEME-TFSI) mixed with lithium bis(trifluoromethanesulfonyl)imide ([Li][TFSI]). For a negative gate voltage ($V_\mathrm{G}$), O$^{2-}$ accumulates on the surface of the film and then enters into the film \cite{Lu2017}, while Li$^{+}$ escapes from the film to the liquid [Fig. \ref{fig1}(b)] \cite{Maruyama2015,Yoshimatsu2015}. To balance the charge, holes are injected into the film. In contrast, applying positive $V_\mathrm{G}$ results in O$^{2-}$ deintercalation \cite{Perez2017,Zhang2017} and Li$^{+}$ intercalation \cite{Lei2017,Ying2018}, which corresponds to electron-doping [Fig. \ref{fig1}(c)]. The amount of doped charge per unit volume is denoted by the quantity $Q$, which is calculated by the ratio of the temporal integral of leak current ($I_\mathrm{G}$) to the sample volume (see Supplemental Material \cite{supplemental} for calculation method). \emph{In situ} XRD measurements are used to identify the structural phase and the lattice constant.

\begin{figure}[ht!]
	\centering
	\includegraphics[width=\linewidth]{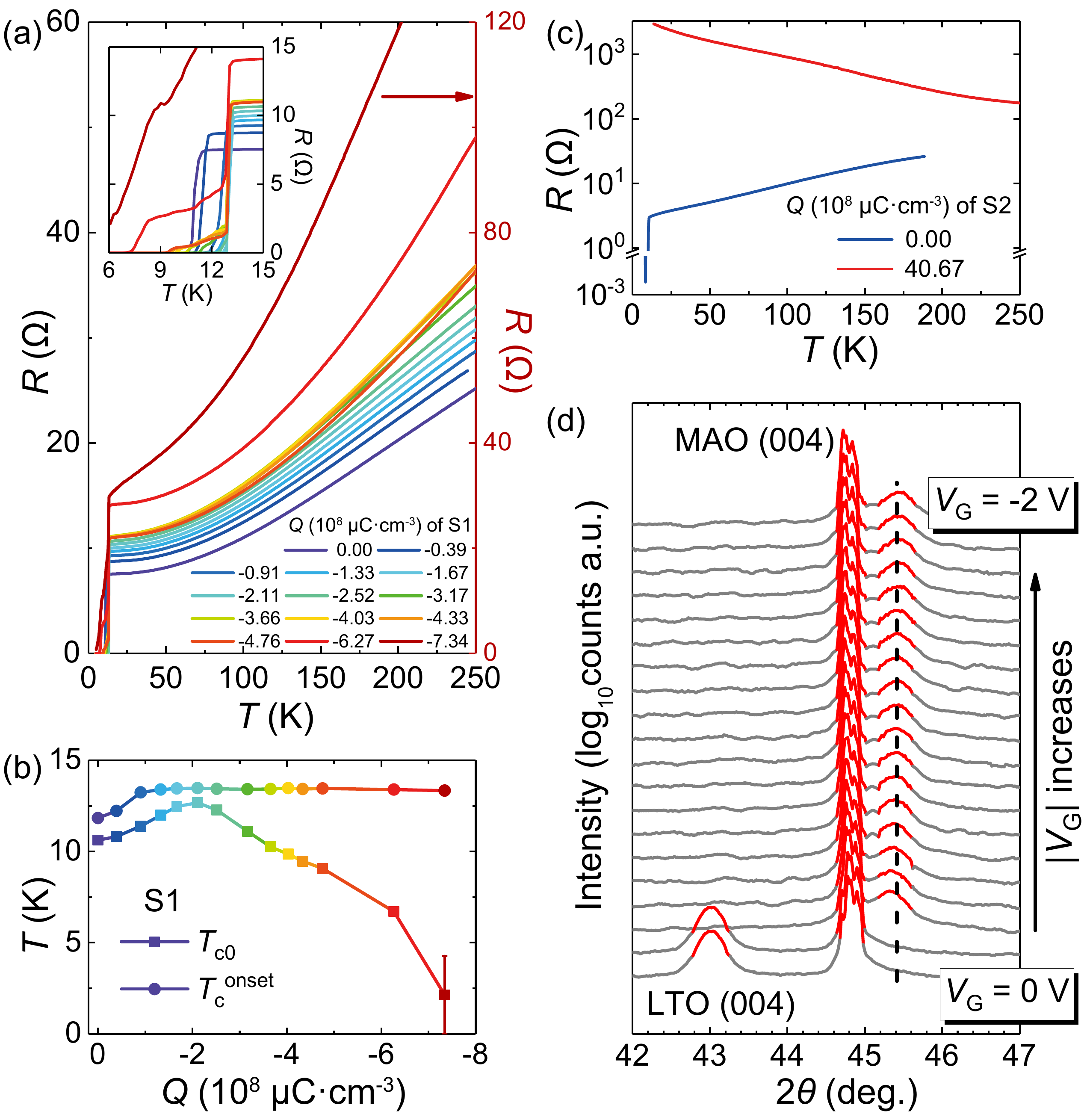}
	\caption{(a) The $R(T)$ curves of sample S1 under various $Q$ with negative $V_\mathrm{G}$. Inset: Zoom-in $R(T)$ curves of S1. (b) The $Q$-dependent $T_\mathrm{c0}$ and $T_\mathrm{c}^\mathrm{onset}$ of S1. (c) The $R(T)$ curves of sample S2 under different $Q$. (d) \emph{In situ} XRD patterns of sample S3. The gating experiment is carried out at room temperature with a $V_\mathrm{G}$ interval of $-0.2$ V. Two rounds of scanning are performed at each $V_\mathrm{G}$ except for initial 0 V (one round) and final $-2$ V (one round). The time interval for each curve is 8.5 minutes.}
	\label{fig2}
\end{figure}

Figure \ref{fig2}(a) shows the $R(T)$ of sample S1 with negative $V_\mathrm{G}$. With the increase of $|Q|$, the normal-state resistance increases monotonically. The inset in Fig. \ref{fig2}(a) exhibits the detailed superconducting transition, and Fig. \ref{fig2}(b) summaries the evolution of $T_\mathrm{c0}$ as well as the onset transition temperature ($T_\mathrm{c}^\mathrm{onset}$). At first, both $T_\mathrm{c0}$ and $T_\mathrm{c}^\mathrm{onset}$ increase, then $T_\mathrm{c0}$ drops to zero while $T_\mathrm{c}^\mathrm{onset}$ remains almost constant. The almost unchanged $T_\mathrm{c}^\mathrm{onset}$, together with the inhomogeneous superconducting transition, may be caused by inhomogeneous gating which is hard to avoid for thick films and has been observed in other systems \cite{Qin2020}. The evolution of $T_\mathrm{c0}$ and $T_\mathrm{c}^\mathrm{onset}$ observed here reconciles the difference in experimental results between ILG and other methods, such as soaking samples in hydrochloric acid and exposing samples to air \cite{Rygula1993,Moshopoulou1993,Moshopoulou1994,Moshopoulou1999,Maruyama2015}.

At a higher doping level where Ti loses almost all valence electrons, a phase transition from superconductor to insulator is expected. Sample S2 has been doped with large amount of holes and the phase transition is achieved [Fig. \ref{fig2}(c)]. When $V_\mathrm{G}$ is removed, the superconductivity of S2 cannot recover, suggesting an irreversible SIT. Such a SIT may be accompanied by a chemical phase change, such as the transition from LTO to insulating Li$_4$Ti$_5$O$_{12}$ when we dope LTO with holes through increasing its O$^{2-}$ content \cite{Jia2018}. Figure \ref{fig2}(d) shows the \emph{in situ} $\theta$-$2\theta$ XRD spectra performed in sample S3 at room temperature. With hole-doping, the $(004)$ diffraction peak of LTO ($c=8.40$ {\AA}) disappears, while a new diffraction peak emerges. The new diffraction peak corresponds to a $c$-axis lattice constant of 7.98 {\AA}, which is much smaller than the lattice constant of Li$_4$Ti$_5$O$_{12}$ (8.36 {\AA}) \cite{DESCHANVRES1971699}. After removing the ionic liquid, the XRD pattern remains unchanged and the resistance of S3 exceeds the range of a multimeter (M$\Omega$), which is consistent with the irreversibility of the SIT observed in transport. Here we provide two reasonable explanations for the formation of new phase: One is the change in Ti-sublattice induced by high Li-deficiency, e.g., partial Ti$^{3+,4+}$ migration from $16d$ to $16c$ and $8a$ sites \cite{Rygula1993,Moshopoulou1993}, which has been confirmed in aging samples; The other is the possible anti-Jahn-Teller distortion of Ti-O octahedra induced by the injected O$^{2-}$, which is similar to the distortion of Cu-O octahedra in La$_2$CuO$_{4+y}$ \cite{Saarela2017}.

\begin{figure}[ht!]
	\centering
	\includegraphics[width=\linewidth]{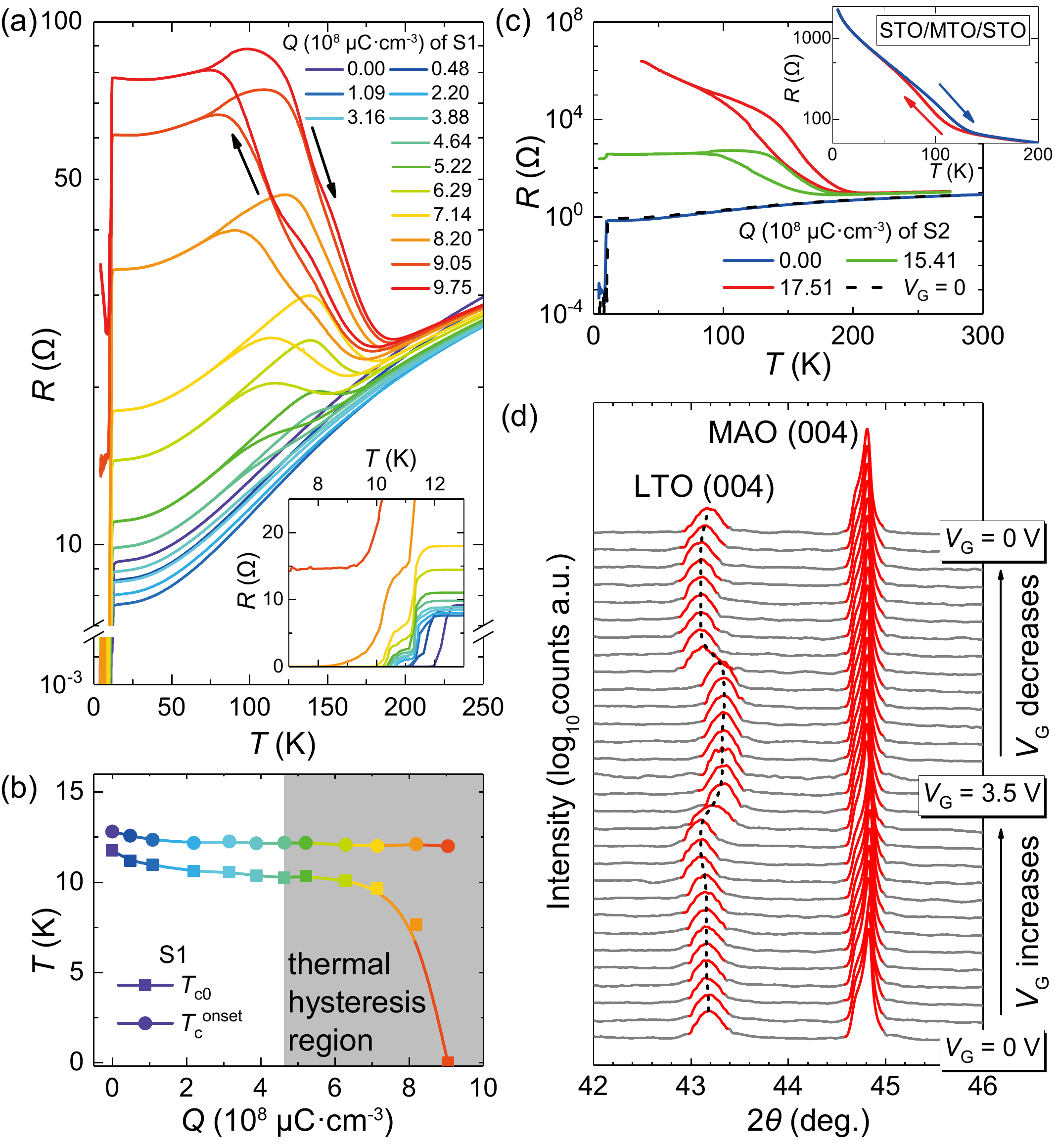}
	\caption{(a) The $R(T)$ curves of sample S1 under various $Q$ with positive $V_\mathrm{G}$. Inset: Zoom-in $R(T)$ curves. (b) The $Q$-dependent $T_\mathrm{c0}$ and $T_\mathrm{c}^\mathrm{onset}$ of S1. (c) The $R(T)$ curves of sample S2 with typical $Q$. The black dash line is the $R(T)$ curve after removing $V_\mathrm{G}$. Inset: The $R(T)$ curves of MTO layer in SrTiO$_3$/MTO/SrTiO$_3$ heterostructure. (d) \emph{In situ} XRD patterns of sample S4. The gating experiment is carried out at room temperature with a $V_\mathrm{G}$ interval of 0.5 V. Two rounds of scanning are performed at each $V_\mathrm{G}$ except for initial 0 V (one round) and 3.5 V (three rounds). The time interval for each curve is 8.5 minutes.}
	\label{fig3}
\end{figure}

The transport measurements of electron-doped S1 and S2 are carried out before samples are doped with holes. The $R(T)$ of S1 with positive $V_\mathrm{G}$ shows many interesting behaviors, as indicated in Fig. \ref{fig3}(a). Different to the case of hole-doping, the resistance of electron-doped S1 in the high temperature region ($>200$ K) changes slightly and keeps positive $\mathrm{d}R/\mathrm{d}T$ during the whole gating process. In the middle temperature region (80 to 200 K), the resistance that originally shows metallic behavior turns to exhibit insulating behavior as $Q$ increases, and a robust thermal hysteresis emerges. The thermal hysteresis is also observed in electron-doped sample S2 and S5, as shown in Fig. \ref{fig3}(c) and Supplemental Material \cite{supplemental}. The appearance of thermal hysteresis implies the occurrence of a first-order phase transition, which is usually accompanied by changes in symmetry of electron charge, spin, and/or orbital. At lower temperature ($<80$ K), LTO regains metallic behavior and finally enters the superconducting state. The inset in Fig. \ref{fig3}(a) shows the detailed superconducting transition and Fig. \ref{fig3}(b) summaries the evolution of $T_\mathrm{c0}$ as well as $T_\mathrm{c}^\mathrm{onset}$. It is obvious that the superconductivity of LTO is suppressed during the electron-doping process. At a higher doping level, as the curve with $Q=9.75\times 10^8\mu\mathrm{C}\cdot\mathrm{cm}^{-3}$ in Fig. \ref{fig3}(a) and the curve with $Q=1.751\times 10^9\mu\mathrm{C}\cdot\mathrm{cm}^{-3}$ in Fig. \ref{fig3}(c) show, a phase transition from superconductor to insulator occurs. This SIT is unexpected because the carrier density increases during the electron-doping process. Although the resistance at low temperature spans more than six orders of magnitude under gating, the superconductivity can recover after removing $V_\mathrm{G}$, as indicated by the black dash line in Fig. \ref{fig3}(c).

In order to clarify the mechanism of the SIT in electron-doped LTO, we perform \emph{in situ} XRD measurements on sample S4 with positive $V_\mathrm{G}$ at room temperature. Figure \ref{fig3}(d) shows the \emph{in situ} $\theta$-$2\theta$ patterns. The $c$-axis lattice constant of S4 is 8.37 {\AA}, which is slightly smaller than that of S3 but still falls in the reasonable range of LTO lattice constant \cite{Jin2015,Jia2018}. When $V_\mathrm{G}$ increases to 3.5 V, the diffraction peak moves to a higher angle, corresponding a decrease of lattice constant from 8.37 to 8.35 {\AA}. Here the lattice constant varies by $0.24\%$, much less than the $5\%$ in the case of hole-doping. Furthermore, the diffraction peak returns to its original position when $V_\mathrm{G}$ decreases, which demonstrates again that the effect of electron-doping is reversible and the mechanisms of SIT observed in electron- and hole-doping regions are different.

The transport and structural results in heavily electron-doped LTO are reminiscent of the so-called ``robust'' insulating spinel oxide MTO with a complicated orbital ordering at low temperature \cite{Schmidt2004}. Nevertheless, superconductivity can be achieved in MTO by suppressing orbital ordering \cite{Hu2020}. Similar to the case of LTO, the lattice constant of the insulating MTO decreases by $0.23\%$ compared with the superconducting MTO \cite{Hu2020}. Besides, the insulating MTO layer, in SrTiO$_3$/MTO/SrTiO$_3$ heterostructure, shows a clear thermal hysteresis in $R(T)$ curves [Inset of Fig. \ref{fig3}(c)], while the superconducting MTO layers exhibit no signs of thermal hysteresis \cite{Hu2020}. Therefore, LTO and MTO are expected to share the same origin of the SIT. It is well accepted that the electron correlations are significant in insulating MTO, given that energy bands split at low temperature due to the orbital ordering \cite{Schmidt2004,Khomskii2005}, and the Hubbard interaction term should be included in theoretical calculations to reproduce the insulating character \cite{Hu2020}. As a result, the SIT in electron-doped LTO likely stems from the enhanced $3d$ electron correlations.

\begin{figure}[ht!]
	\centering
	\includegraphics[width=\linewidth]{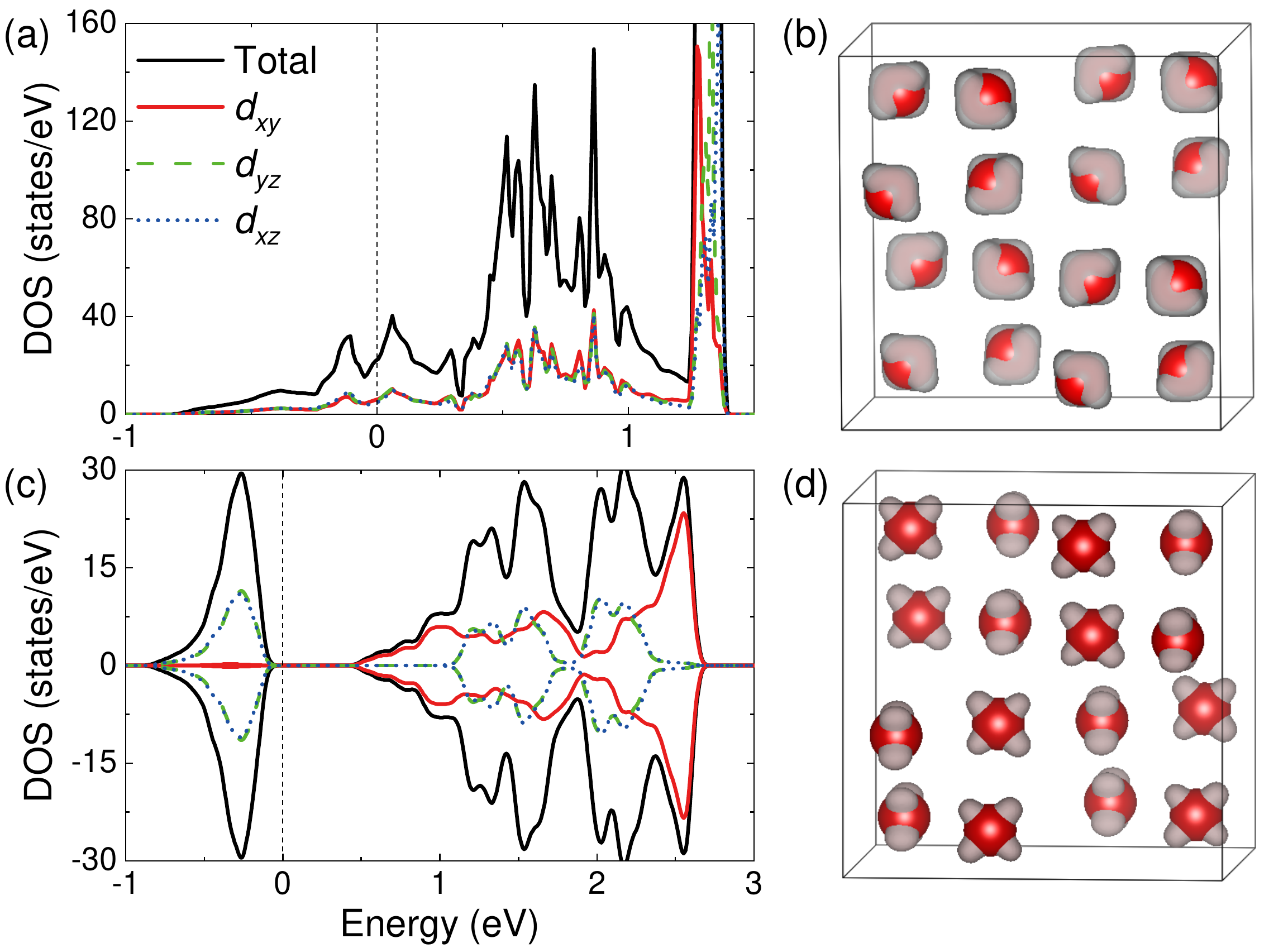}
	\caption{(a,c) Calculated density of states for undoped (a) and electron-doped (c) LTO. (b,d) The integrated charge densities for undoped (b) and electron-doped LTO (d). For electron-doped LTO, the doping concentration is set $0.5e/\mathrm{Ti}$ and a Hubbard $U$ of 2.5 eV is included.}
	\label{fig4}
\end{figure}

In order to further confirm the role of electron correlations in the SIT, first-principles electronic structure calculations have been performed (see Supplemental Material \cite{supplemental} for computational details). It is found that the undoped LTO tends to relax to the nonmagnetic state with equilong Ti-Ti distances \cite{Akimoto1992}, even if the initial state is set to a magnetic singlet state with dimerized Ti-Ti bonds as in MTO \cite{Schmidt2004,ZHU2007578}. The calculated finite electronic states at the Fermi level, as Fig. \ref{fig4}(a) shows, indicates the metallic character and is consistent with previous theoretical study \cite{Satpathy1987}. The $t_\mathrm{2g}$ orbitals of Ti around the Fermi level is found nearly degenerate without showing up the orbital ordering [Fig. \ref{fig4}(b)]. Once enough electrons are introduced into LTO, dramatic changes take place. At the doping concentration of $0.5e/\mathrm{Ti}$, the nominal valence of Ti in doped LTO is equivalent to that in MTO, thus a Hubbard $U$ of 2.5 eV as in MTO \cite{Hu2020} is included to describe the correlation effect among Ti $3d$ electrons. In this case, the magnetic singlet state with dimerized Ti-Ti bonds is more stable than the nonmagnetic state. Figure \ref{fig4}(c) displays the corresponding density of states for the doped LTO in the magnetic singlet state. The finite band gap indicates the metal-insulator transition in electron-doped LTO, which is consistent with our resistance measurements [Fig. \ref{fig3}(c)]. Moreover, the integrated charge densities near the Fermi level show clear orbital ordering in the Ti sublattice [Fig. \ref{fig4}(d)], reproducing the scenario of MTO \cite{Leoni2008,Hu2020}. It is worth noting that the electron-doped LTO is always a nonmagnetic metal without the inclusion of Hubbard $U$ (see Supplemental Material \cite{supplemental}, Fig. S2). These calculation results suggest that electron-doping in LTO enhances the electron correlations, which favors the orbital ordering accompanied with the SIT.

Finally, we discuss the origin of the thermal hysteresis observed in electron-doped LTO. On microscopic scale, the appearance of thermal hysteresis indicates the formation of clusters with symmetry breaking \cite{Zhang2002,Sipos2008,Mattoni2016}. The symmetry breaking in clusters is usually associated with the change in degrees of freedom of electron orbital \cite{Haverkort2005,Zhou2006,Liang2020}, spin \cite{Zhang2002,Catalano2015}, and charge \cite{Leininger2011,Yu2015}. Previous works such as crystal structure \cite{Isobe2002,Schmidt2004} and optical measurements \cite{Zhou2006} have shown that the appearance of thermal hysteresis in MTO bulk corresponds to the formation of orbital ordering. Since LTO and MTO have many common features, the thermal hysteresis in LTO may also be attributed to the orbital ordering. This inference is reasonable because the pristine LTO has orbital-related state below 50 K \cite{Jin2015}, and our calculations support the appearance of orbital ordering in electron-doped LTO [Fig. \ref{fig4}(d)]. We note $T_\mathrm{c0}$ of electron-doped LTO drops dramatically after thermal hysteresis emerges [Fig. \ref{fig3}(b)], indicating a possible competition between superconductivity and orbital ordering. In addition, we analyze the temperature dependence of resistance of electron-doped S1 by the function $R = R_0 + A T^{\alpha}$, where $R_0$, $A$, and $\alpha$ are the residual resistance, a constant, and the exponent index that is sensitive to the orbital character of electrons, respectively \cite{Piatti2018,Zhou2020} (see Supplemental Material \cite{supplemental} for detailed analysis). When the thermal hysteresis emerges, the $\alpha$ increases with the increase of $Q$, suggesting that the orbital character of electrons in the percolative metallic/superconducting paths is also affected.

In conclusion, we perform systematic transport and \emph{in situ} XRD measurements of LTO films based on ILG, and obtain a detailed phase diagram. Two superconductor-insulator phase transitions are found at both high electron- and hole- doping level. Combining the experimental results and theoretical calculations, we argue that the irreversible SIT in hole-doping region is triggered by the decrease of valence electrons, while the reversible SIT in electron-doping region is caused by the enhanced $d$-$d$ electron correlations. The thermal hysteresis in electron-doped LTO likely stems from orbital ordering. The detailed interplay between superconductivity and orbital ordering deserves further study, which not only complements the effects of $3d$ electrons behavior on superconductivity but also helps to understand the orbital-selective pairing in iron-based superconductors \cite{Kreisel2017,*Sprau2017}.

\begin{acknowledgements}
This work was supported by the National Key R\& D Program of China (Grants No. 2016YFA0300301, No. 2017YFA0302900, No. 2017YFA0303003, No. 2018YFB0704102, No. 2018YFA0305604, and No. 2019YFA0308402), the National Natural Science Foundation of China (Grants No. 11674374, No. 11774424, No. 11834016, No. 11804378, No. 11961141008, and No. 11927808), the Strategic Priority Research Program (B) of Chinese Academy of Sciences (No. XDB25000000), the Key Research Program of Frontier Sciences, CAS (Grants No. QYZDB-SSW-SLH008, No. QYZDY-SSW-SLH001, and No. QYZDB-SSW-JSC035), CAS Interdisciplinary Innovation Team, Beijing Natural Science Foundation (Grants No. Z190008 and No. Z200005), the Fundamental Research Funds for the Central Universities, and the Research Funds of Renmin University of China (No. 19XNLG13). Computational resources are provided by the Physical Laboratory of High Performance Computing at Renmin University of China. G.H. thanks the Alexander von Humboldt Foundation for support from a research fellowship.
\end{acknowledgements}

\bibliography{refv} 

\begin{thebibliography}{59}%
\makeatletter
\providecommand \@ifxundefined [1]{%
 \@ifx{#1\undefined}
}%
\providecommand \@ifnum [1]{%
 \ifnum #1\expandafter \@firstoftwo
 \else \expandafter \@secondoftwo
 \fi
}%
\providecommand \@ifx [1]{%
 \ifx #1\expandafter \@firstoftwo
 \else \expandafter \@secondoftwo
 \fi
}%
\providecommand \natexlab [1]{#1}%
\providecommand \enquote  [1]{``#1''}%
\providecommand \bibnamefont  [1]{#1}%
\providecommand \bibfnamefont [1]{#1}%
\providecommand \citenamefont [1]{#1}%
\providecommand \href@noop [0]{\@secondoftwo}%
\providecommand \href [0]{\begingroup \@sanitize@url \@href}%
\providecommand \@href[1]{\@@startlink{#1}\@@href}%
\providecommand \@@href[1]{\endgroup#1\@@endlink}%
\providecommand \@sanitize@url [0]{\catcode `\\12\catcode `\$12\catcode
  `\&12\catcode `\#12\catcode `\^12\catcode `\_12\catcode `\%12\relax}%
\providecommand \@@startlink[1]{}%
\providecommand \@@endlink[0]{}%
\providecommand \url  [0]{\begingroup\@sanitize@url \@url }%
\providecommand \@url [1]{\endgroup\@href {#1}{\urlprefix }}%
\providecommand \urlprefix  [0]{URL }%
\providecommand \Eprint [0]{\href }%
\providecommand \doibase [0]{https://doi.org/}%
\providecommand \selectlanguage [0]{\@gobble}%
\providecommand \bibinfo  [0]{\@secondoftwo}%
\providecommand \bibfield  [0]{\@secondoftwo}%
\providecommand \translation [1]{[#1]}%
\providecommand \BibitemOpen [0]{}%
\providecommand \bibitemStop [0]{}%
\providecommand \bibitemNoStop [0]{.\EOS\space}%
\providecommand \EOS [0]{\spacefactor3000\relax}%
\providecommand \BibitemShut  [1]{\csname bibitem#1\endcsname}%
\let\auto@bib@innerbib\@empty
\bibitem [{\citenamefont {Paglione}\ and\ \citenamefont
  {Greene}(2010)}]{Paglione2010}%
  \BibitemOpen
  \bibfield  {author} {\bibinfo {author} {\bibfnamefont {J.}~\bibnamefont
  {Paglione}}\ and\ \bibinfo {author} {\bibfnamefont {R.~L.}\ \bibnamefont
  {Greene}},\ }\href {https://doi.org/10.1038/nphys1759} {\bibfield  {journal}
  {\bibinfo  {journal} {Nat. Phys.}\ }\textbf {\bibinfo {volume} {6}},\
  \bibinfo {pages} {645} (\bibinfo {year} {2010})}\BibitemShut {NoStop}%
\bibitem [{\citenamefont {Jin}\ \emph {et~al.}(2011)\citenamefont {Jin},
  \citenamefont {Butch}, \citenamefont {Kirshenbaum}, \citenamefont
  {Paglione},\ and\ \citenamefont {Greene}}]{Jin2011}%
  \BibitemOpen
  \bibfield  {author} {\bibinfo {author} {\bibfnamefont {K.}~\bibnamefont
  {Jin}}, \bibinfo {author} {\bibfnamefont {N.~P.}\ \bibnamefont {Butch}},
  \bibinfo {author} {\bibfnamefont {K.}~\bibnamefont {Kirshenbaum}}, \bibinfo
  {author} {\bibfnamefont {J.}~\bibnamefont {Paglione}},\ and\ \bibinfo
  {author} {\bibfnamefont {R.~L.}\ \bibnamefont {Greene}},\ }\href
  {https://doi.org/10.1038/nature10308} {\bibfield  {journal} {\bibinfo
  {journal} {Nature (London)}\ }\textbf {\bibinfo {volume} {476}},\ \bibinfo
  {pages} {73} (\bibinfo {year} {2011})}\BibitemShut {NoStop}%
\bibitem [{\citenamefont {Stewart}(2011)}]{Stewart2011}%
  \BibitemOpen
  \bibfield  {author} {\bibinfo {author} {\bibfnamefont {G.~R.}\ \bibnamefont
  {Stewart}},\ }\href {https://doi.org/10.1103/RevModPhys.83.1589} {\bibfield
  {journal} {\bibinfo  {journal} {Rev. Mod. Phys.}\ }\textbf {\bibinfo {volume}
  {83}},\ \bibinfo {pages} {1589} (\bibinfo {year} {2011})}\BibitemShut
  {NoStop}%
\bibitem [{\citenamefont {Keimer}\ \emph {et~al.}(2015)\citenamefont {Keimer},
  \citenamefont {Kivelson}, \citenamefont {Norman}, \citenamefont {Uchida},\
  and\ \citenamefont {Zaanen}}]{Keimer2015}%
  \BibitemOpen
  \bibfield  {author} {\bibinfo {author} {\bibfnamefont {B.}~\bibnamefont
  {Keimer}}, \bibinfo {author} {\bibfnamefont {S.~A.}\ \bibnamefont
  {Kivelson}}, \bibinfo {author} {\bibfnamefont {M.~R.}\ \bibnamefont
  {Norman}}, \bibinfo {author} {\bibfnamefont {S.}~\bibnamefont {Uchida}},\
  and\ \bibinfo {author} {\bibfnamefont {J.}~\bibnamefont {Zaanen}},\ }\href
  {https://doi.org/10.1038/nature14165} {\bibfield  {journal} {\bibinfo
  {journal} {Nature (London)}\ }\textbf {\bibinfo {volume} {518}},\ \bibinfo
  {pages} {179} (\bibinfo {year} {2015})}\BibitemShut {NoStop}%
\bibitem [{\citenamefont {Tunstall}\ \emph {et~al.}(1994)\citenamefont
  {Tunstall}, \citenamefont {Todd}, \citenamefont {Arumugam}, \citenamefont
  {Dai}, \citenamefont {Dalton},\ and\ \citenamefont {Edwards}}]{Tunstall1994}%
  \BibitemOpen
  \bibfield  {author} {\bibinfo {author} {\bibfnamefont {D.~P.}\ \bibnamefont
  {Tunstall}}, \bibinfo {author} {\bibfnamefont {J.~R.~M.}\ \bibnamefont
  {Todd}}, \bibinfo {author} {\bibfnamefont {S.}~\bibnamefont {Arumugam}},
  \bibinfo {author} {\bibfnamefont {G.}~\bibnamefont {Dai}}, \bibinfo {author}
  {\bibfnamefont {M.}~\bibnamefont {Dalton}},\ and\ \bibinfo {author}
  {\bibfnamefont {P.~P.}\ \bibnamefont {Edwards}},\ }\href
  {https://doi.org/10.1103/PhysRevB.50.16541} {\bibfield  {journal} {\bibinfo
  {journal} {Phys. Rev. B}\ }\textbf {\bibinfo {volume} {50}},\ \bibinfo
  {pages} {16541} (\bibinfo {year} {1994})}\BibitemShut {NoStop}%
\bibitem [{\citenamefont {Chen}\ \emph {et~al.}(2011)\citenamefont {Chen},
  \citenamefont {Dong}, \citenamefont {Asokan}, \citenamefont {Chen},
  \citenamefont {Liu}, \citenamefont {Guo}, \citenamefont {Yang}, \citenamefont
  {Chen}, \citenamefont {Hsu}, \citenamefont {Chang},\ and\ \citenamefont
  {Wu}}]{Chen2011}%
  \BibitemOpen
  \bibfield  {author} {\bibinfo {author} {\bibfnamefont {C.~L.}\ \bibnamefont
  {Chen}}, \bibinfo {author} {\bibfnamefont {C.~L.}\ \bibnamefont {Dong}},
  \bibinfo {author} {\bibfnamefont {K.}~\bibnamefont {Asokan}}, \bibinfo
  {author} {\bibfnamefont {J.~L.}\ \bibnamefont {Chen}}, \bibinfo {author}
  {\bibfnamefont {Y.~S.}\ \bibnamefont {Liu}}, \bibinfo {author} {\bibfnamefont
  {J.-H.}\ \bibnamefont {Guo}}, \bibinfo {author} {\bibfnamefont {W.~L.}\
  \bibnamefont {Yang}}, \bibinfo {author} {\bibfnamefont {Y.~Y.}\ \bibnamefont
  {Chen}}, \bibinfo {author} {\bibfnamefont {F.~C.}\ \bibnamefont {Hsu}},
  \bibinfo {author} {\bibfnamefont {C.~L.}\ \bibnamefont {Chang}},\ and\
  \bibinfo {author} {\bibfnamefont {M.~K.}\ \bibnamefont {Wu}},\ }\href
  {https://doi.org/10.1088/0953-2048/24/11/115007} {\bibfield  {journal}
  {\bibinfo  {journal} {Supercond. Sci. Technol.}\ }\textbf {\bibinfo {volume}
  {24}},\ \bibinfo {pages} {115007} (\bibinfo {year} {2011})}\BibitemShut
  {NoStop}%
\bibitem [{\citenamefont {Johnston}(1976)}]{Johnston1976}%
  \BibitemOpen
  \bibfield  {author} {\bibinfo {author} {\bibfnamefont {D.~C.}\ \bibnamefont
  {Johnston}},\ }\href {https://doi.org/10.1007/Bf00654827} {\bibfield
  {journal} {\bibinfo  {journal} {J. Low Temp. Phys.}\ }\textbf {\bibinfo
  {volume} {25}},\ \bibinfo {pages} {145} (\bibinfo {year} {1976})}\BibitemShut
  {NoStop}%
\bibitem [{\citenamefont {Green}\ \emph {et~al.}(1997)\citenamefont {Green},
  \citenamefont {Dalton}, \citenamefont {Prassides}, \citenamefont {Day},\ and\
  \citenamefont {Neumann}}]{Green1997}%
  \BibitemOpen
  \bibfield  {author} {\bibinfo {author} {\bibfnamefont {M.~A.}\ \bibnamefont
  {Green}}, \bibinfo {author} {\bibfnamefont {M.}~\bibnamefont {Dalton}},
  \bibinfo {author} {\bibfnamefont {K.}~\bibnamefont {Prassides}}, \bibinfo
  {author} {\bibfnamefont {P.}~\bibnamefont {Day}},\ and\ \bibinfo {author}
  {\bibfnamefont {D.~A.}\ \bibnamefont {Neumann}},\ }\href
  {https://doi.org/10.1088/0953-8984/9/49/006} {\bibfield  {journal} {\bibinfo
  {journal} {J. Phys.: Condens. Matter}\ }\textbf {\bibinfo {volume} {9}},\
  \bibinfo {pages} {10855} (\bibinfo {year} {1997})}\BibitemShut {NoStop}%
\bibitem [{\citenamefont {Sun}\ \emph {et~al.}(2004)\citenamefont {Sun},
  \citenamefont {Lin}, \citenamefont {Mollah}, \citenamefont {Ho},
  \citenamefont {Yang}, \citenamefont {Hsu}, \citenamefont {Liao},\ and\
  \citenamefont {Wu}}]{Sun2004}%
  \BibitemOpen
  \bibfield  {author} {\bibinfo {author} {\bibfnamefont {C.~P.}\ \bibnamefont
  {Sun}}, \bibinfo {author} {\bibfnamefont {J.-Y.}\ \bibnamefont {Lin}},
  \bibinfo {author} {\bibfnamefont {S.}~\bibnamefont {Mollah}}, \bibinfo
  {author} {\bibfnamefont {P.~L.}\ \bibnamefont {Ho}}, \bibinfo {author}
  {\bibfnamefont {H.~D.}\ \bibnamefont {Yang}}, \bibinfo {author}
  {\bibfnamefont {F.~C.}\ \bibnamefont {Hsu}}, \bibinfo {author} {\bibfnamefont
  {Y.~C.}\ \bibnamefont {Liao}},\ and\ \bibinfo {author} {\bibfnamefont
  {M.~K.}\ \bibnamefont {Wu}},\ }\href
  {https://doi.org/10.1103/PhysRevB.70.054519} {\bibfield  {journal} {\bibinfo
  {journal} {Phys. Rev. B}\ }\textbf {\bibinfo {volume} {70}},\ \bibinfo
  {pages} {054519} (\bibinfo {year} {2004})}\BibitemShut {NoStop}%
\bibitem [{\citenamefont {Tang}\ \emph {et~al.}(2006)\citenamefont {Tang},
  \citenamefont {Zou}, \citenamefont {Shan}, \citenamefont {Dong},
  \citenamefont {Che},\ and\ \citenamefont {Wen}}]{Tang2006}%
  \BibitemOpen
  \bibfield  {author} {\bibinfo {author} {\bibfnamefont {L.}~\bibnamefont
  {Tang}}, \bibinfo {author} {\bibfnamefont {P.~Y.}\ \bibnamefont {Zou}},
  \bibinfo {author} {\bibfnamefont {L.}~\bibnamefont {Shan}}, \bibinfo {author}
  {\bibfnamefont {A.~F.}\ \bibnamefont {Dong}}, \bibinfo {author}
  {\bibfnamefont {G.~C.}\ \bibnamefont {Che}},\ and\ \bibinfo {author}
  {\bibfnamefont {H.~H.}\ \bibnamefont {Wen}},\ }\href
  {https://doi.org/10.1103/PhysRevB.73.184521} {\bibfield  {journal} {\bibinfo
  {journal} {Phys. Rev. B}\ }\textbf {\bibinfo {volume} {73}},\ \bibinfo
  {pages} {184521} (\bibinfo {year} {2006})}\BibitemShut {NoStop}%
\bibitem [{\citenamefont {Edwards}\ \emph {et~al.}(1984)\citenamefont
  {Edwards}, \citenamefont {Egdell}, \citenamefont {Fragala}, \citenamefont
  {Goodenough}, \citenamefont {Harrison}, \citenamefont {Orchard},\ and\
  \citenamefont {Scott}}]{Edwards1984}%
  \BibitemOpen
  \bibfield  {author} {\bibinfo {author} {\bibfnamefont {P.}~\bibnamefont
  {Edwards}}, \bibinfo {author} {\bibfnamefont {R.}~\bibnamefont {Egdell}},
  \bibinfo {author} {\bibfnamefont {I.}~\bibnamefont {Fragala}}, \bibinfo
  {author} {\bibfnamefont {J.}~\bibnamefont {Goodenough}}, \bibinfo {author}
  {\bibfnamefont {M.}~\bibnamefont {Harrison}}, \bibinfo {author}
  {\bibfnamefont {A.}~\bibnamefont {Orchard}},\ and\ \bibinfo {author}
  {\bibfnamefont {E.}~\bibnamefont {Scott}},\ }\href
  {https://doi.org/doi.org/10.1016/0022-4596(84)90140-3} {\bibfield  {journal}
  {\bibinfo  {journal} {J. Solid State Chem.}\ }\textbf {\bibinfo {volume}
  {54}},\ \bibinfo {pages} {127 } (\bibinfo {year} {1984})}\BibitemShut
  {NoStop}%
\bibitem [{\citenamefont {Anderson}(1987)}]{Anderson1987}%
  \BibitemOpen
  \bibfield  {author} {\bibinfo {author} {\bibfnamefont {P.~W.}\ \bibnamefont
  {Anderson}},\ }\href {https://doi.org/10.1126/science.235.4793.1196}
  {\bibfield  {journal} {\bibinfo  {journal} {Science}\ }\textbf {\bibinfo
  {volume} {235}},\ \bibinfo {pages} {1196} (\bibinfo {year}
  {1987})}\BibitemShut {NoStop}%
\bibitem [{\citenamefont {Anderson}\ \emph {et~al.}(1987)\citenamefont
  {Anderson}, \citenamefont {Baskaran}, \citenamefont {Zou},\ and\
  \citenamefont {Hsu}}]{Anderson1987_2}%
  \BibitemOpen
  \bibfield  {author} {\bibinfo {author} {\bibfnamefont {P.~W.}\ \bibnamefont
  {Anderson}}, \bibinfo {author} {\bibfnamefont {G.}~\bibnamefont {Baskaran}},
  \bibinfo {author} {\bibfnamefont {Z.}~\bibnamefont {Zou}},\ and\ \bibinfo
  {author} {\bibfnamefont {T.}~\bibnamefont {Hsu}},\ }\href
  {https://doi.org/10.1103/PhysRevLett.58.2790} {\bibfield  {journal} {\bibinfo
   {journal} {Phys. Rev. Lett.}\ }\textbf {\bibinfo {volume} {58}},\ \bibinfo
  {pages} {2790} (\bibinfo {year} {1987})}\BibitemShut {NoStop}%
\bibitem [{\citenamefont {Jin}\ \emph {et~al.}(2015)\citenamefont {Jin},
  \citenamefont {He}, \citenamefont {Zhang}, \citenamefont {Maruyama},
  \citenamefont {Yasui}, \citenamefont {Suchoski}, \citenamefont {Shin},
  \citenamefont {Jiang}, \citenamefont {Yu}, \citenamefont {Yuan},
  \citenamefont {Shan}, \citenamefont {Kusmartsev}, \citenamefont {Greene},\
  and\ \citenamefont {Takeuchi}}]{Jin2015}%
  \BibitemOpen
  \bibfield  {author} {\bibinfo {author} {\bibfnamefont {K.}~\bibnamefont
  {Jin}}, \bibinfo {author} {\bibfnamefont {G.}~\bibnamefont {He}}, \bibinfo
  {author} {\bibfnamefont {X.}~\bibnamefont {Zhang}}, \bibinfo {author}
  {\bibfnamefont {S.}~\bibnamefont {Maruyama}}, \bibinfo {author}
  {\bibfnamefont {S.}~\bibnamefont {Yasui}}, \bibinfo {author} {\bibfnamefont
  {R.}~\bibnamefont {Suchoski}}, \bibinfo {author} {\bibfnamefont
  {J.}~\bibnamefont {Shin}}, \bibinfo {author} {\bibfnamefont {Y.}~\bibnamefont
  {Jiang}}, \bibinfo {author} {\bibfnamefont {H.~S.}\ \bibnamefont {Yu}},
  \bibinfo {author} {\bibfnamefont {J.}~\bibnamefont {Yuan}}, \bibinfo {author}
  {\bibfnamefont {L.}~\bibnamefont {Shan}}, \bibinfo {author} {\bibfnamefont
  {F.~V.}\ \bibnamefont {Kusmartsev}}, \bibinfo {author} {\bibfnamefont
  {R.~L.}\ \bibnamefont {Greene}},\ and\ \bibinfo {author} {\bibfnamefont
  {I.}~\bibnamefont {Takeuchi}},\ }\href {https://doi.org/10.1038/ncomms8183}
  {\bibfield  {journal} {\bibinfo  {journal} {Nat. Commun.}\ }\textbf {\bibinfo
  {volume} {6}},\ \bibinfo {pages} {7183} (\bibinfo {year} {2015})}\BibitemShut
  {NoStop}%
\bibitem [{\citenamefont {Chen}\ \emph {et~al.}(2017)\citenamefont {Chen},
  \citenamefont {Jia}, \citenamefont {Zhang}, \citenamefont {Fang},
  \citenamefont {Jin}, \citenamefont {Richard},\ and\ \citenamefont
  {Ding}}]{Chen2017}%
  \BibitemOpen
  \bibfield  {author} {\bibinfo {author} {\bibfnamefont {D.}~\bibnamefont
  {Chen}}, \bibinfo {author} {\bibfnamefont {Y.-L.}\ \bibnamefont {Jia}},
  \bibinfo {author} {\bibfnamefont {T.-T.}\ \bibnamefont {Zhang}}, \bibinfo
  {author} {\bibfnamefont {Z.}~\bibnamefont {Fang}}, \bibinfo {author}
  {\bibfnamefont {K.}~\bibnamefont {Jin}}, \bibinfo {author} {\bibfnamefont
  {P.}~\bibnamefont {Richard}},\ and\ \bibinfo {author} {\bibfnamefont
  {H.}~\bibnamefont {Ding}},\ }\href
  {https://doi.org/10.1103/PhysRevB.96.094501} {\bibfield  {journal} {\bibinfo
  {journal} {Phys. Rev. B}\ }\textbf {\bibinfo {volume} {96}},\ \bibinfo
  {pages} {094501} (\bibinfo {year} {2017})}\BibitemShut {NoStop}%
\bibitem [{\citenamefont {Okada}\ \emph {et~al.}(2017)\citenamefont {Okada},
  \citenamefont {Ando}, \citenamefont {Shimizu}, \citenamefont {Minamitani},
  \citenamefont {Shiraki}, \citenamefont {Watanabe},\ and\ \citenamefont
  {Hitosugi}}]{Okada2017}%
  \BibitemOpen
  \bibfield  {author} {\bibinfo {author} {\bibfnamefont {Y.}~\bibnamefont
  {Okada}}, \bibinfo {author} {\bibfnamefont {Y.}~\bibnamefont {Ando}},
  \bibinfo {author} {\bibfnamefont {R.}~\bibnamefont {Shimizu}}, \bibinfo
  {author} {\bibfnamefont {E.}~\bibnamefont {Minamitani}}, \bibinfo {author}
  {\bibfnamefont {S.}~\bibnamefont {Shiraki}}, \bibinfo {author} {\bibfnamefont
  {S.}~\bibnamefont {Watanabe}},\ and\ \bibinfo {author} {\bibfnamefont
  {T.}~\bibnamefont {Hitosugi}},\ }\href {https://doi.org/10.1038/ncomms15975}
  {\bibfield  {journal} {\bibinfo  {journal} {Nat. Commun.}\ }\textbf {\bibinfo
  {volume} {8}},\ \bibinfo {pages} {15975} (\bibinfo {year}
  {2017})}\BibitemShut {NoStop}%
\bibitem [{\citenamefont {He}\ \emph {et~al.}(2017)\citenamefont {He},
  \citenamefont {Jia}, \citenamefont {Hou}, \citenamefont {Wei}, \citenamefont
  {Xie}, \citenamefont {Yang}, \citenamefont {Shi}, \citenamefont {Yuan},
  \citenamefont {Shan}, \citenamefont {Zhu}, \citenamefont {Li}, \citenamefont
  {Gu}, \citenamefont {Liu}, \citenamefont {Xiang},\ and\ \citenamefont
  {Jin}}]{He2017}%
  \BibitemOpen
  \bibfield  {author} {\bibinfo {author} {\bibfnamefont {G.}~\bibnamefont
  {He}}, \bibinfo {author} {\bibfnamefont {Y.}~\bibnamefont {Jia}}, \bibinfo
  {author} {\bibfnamefont {X.}~\bibnamefont {Hou}}, \bibinfo {author}
  {\bibfnamefont {Z.}~\bibnamefont {Wei}}, \bibinfo {author} {\bibfnamefont
  {H.}~\bibnamefont {Xie}}, \bibinfo {author} {\bibfnamefont {Z.}~\bibnamefont
  {Yang}}, \bibinfo {author} {\bibfnamefont {J.}~\bibnamefont {Shi}}, \bibinfo
  {author} {\bibfnamefont {J.}~\bibnamefont {Yuan}}, \bibinfo {author}
  {\bibfnamefont {L.}~\bibnamefont {Shan}}, \bibinfo {author} {\bibfnamefont
  {B.}~\bibnamefont {Zhu}}, \bibinfo {author} {\bibfnamefont {H.}~\bibnamefont
  {Li}}, \bibinfo {author} {\bibfnamefont {L.}~\bibnamefont {Gu}}, \bibinfo
  {author} {\bibfnamefont {K.}~\bibnamefont {Liu}}, \bibinfo {author}
  {\bibfnamefont {T.}~\bibnamefont {Xiang}},\ and\ \bibinfo {author}
  {\bibfnamefont {K.}~\bibnamefont {Jin}},\ }\href
  {https://doi.org/10.1103/PhysRevB.95.054510} {\bibfield  {journal} {\bibinfo
  {journal} {Phys. Rev. B}\ }\textbf {\bibinfo {volume} {95}},\ \bibinfo
  {pages} {054510} (\bibinfo {year} {2017})}\BibitemShut {NoStop}%
\bibitem [{\citenamefont {Wei}\ \emph {et~al.}(2019)\citenamefont {Wei},
  \citenamefont {He}, \citenamefont {Hu}, \citenamefont {Feng}, \citenamefont
  {Wei}, \citenamefont {Ho}, \citenamefont {Li}, \citenamefont {Yuan},
  \citenamefont {Xi}, \citenamefont {Wang}, \citenamefont {Chen}, \citenamefont
  {Zhu}, \citenamefont {Zhou}, \citenamefont {Dong}, \citenamefont {Pi},
  \citenamefont {Kusmartseva}, \citenamefont {Kusmartsev}, \citenamefont
  {Zhao},\ and\ \citenamefont {Jin}}]{PhysRevB.100.184509}%
  \BibitemOpen
  \bibfield  {author} {\bibinfo {author} {\bibfnamefont {Z.}~\bibnamefont
  {Wei}}, \bibinfo {author} {\bibfnamefont {G.}~\bibnamefont {He}}, \bibinfo
  {author} {\bibfnamefont {W.}~\bibnamefont {Hu}}, \bibinfo {author}
  {\bibfnamefont {Z.}~\bibnamefont {Feng}}, \bibinfo {author} {\bibfnamefont
  {X.}~\bibnamefont {Wei}}, \bibinfo {author} {\bibfnamefont {C.~Y.}\
  \bibnamefont {Ho}}, \bibinfo {author} {\bibfnamefont {Q.}~\bibnamefont {Li}},
  \bibinfo {author} {\bibfnamefont {J.}~\bibnamefont {Yuan}}, \bibinfo {author}
  {\bibfnamefont {C.}~\bibnamefont {Xi}}, \bibinfo {author} {\bibfnamefont
  {Z.}~\bibnamefont {Wang}}, \bibinfo {author} {\bibfnamefont {Q.}~\bibnamefont
  {Chen}}, \bibinfo {author} {\bibfnamefont {B.}~\bibnamefont {Zhu}}, \bibinfo
  {author} {\bibfnamefont {F.}~\bibnamefont {Zhou}}, \bibinfo {author}
  {\bibfnamefont {X.}~\bibnamefont {Dong}}, \bibinfo {author} {\bibfnamefont
  {L.}~\bibnamefont {Pi}}, \bibinfo {author} {\bibfnamefont {A.}~\bibnamefont
  {Kusmartseva}}, \bibinfo {author} {\bibfnamefont {F.~V.}\ \bibnamefont
  {Kusmartsev}}, \bibinfo {author} {\bibfnamefont {Z.}~\bibnamefont {Zhao}},\
  and\ \bibinfo {author} {\bibfnamefont {K.}~\bibnamefont {Jin}},\ }\href
  {https://doi.org/10.1103/PhysRevB.100.184509} {\bibfield  {journal} {\bibinfo
   {journal} {Phys. Rev. B}\ }\textbf {\bibinfo {volume} {100}},\ \bibinfo
  {pages} {184509} (\bibinfo {year} {2019})}\BibitemShut {NoStop}%
\bibitem [{\citenamefont {Ye}\ \emph {et~al.}(2012)\citenamefont {Ye},
  \citenamefont {Zhang}, \citenamefont {Akashi}, \citenamefont {Bahramy},
  \citenamefont {Arita},\ and\ \citenamefont {Iwasa}}]{Ye1193}%
  \BibitemOpen
  \bibfield  {author} {\bibinfo {author} {\bibfnamefont {J.~T.}\ \bibnamefont
  {Ye}}, \bibinfo {author} {\bibfnamefont {Y.~J.}\ \bibnamefont {Zhang}},
  \bibinfo {author} {\bibfnamefont {R.}~\bibnamefont {Akashi}}, \bibinfo
  {author} {\bibfnamefont {M.~S.}\ \bibnamefont {Bahramy}}, \bibinfo {author}
  {\bibfnamefont {R.}~\bibnamefont {Arita}},\ and\ \bibinfo {author}
  {\bibfnamefont {Y.}~\bibnamefont {Iwasa}},\ }\href
  {https://doi.org/10.1126/science.1228006} {\bibfield  {journal} {\bibinfo
  {journal} {Science}\ }\textbf {\bibinfo {volume} {338}},\ \bibinfo {pages}
  {1193} (\bibinfo {year} {2012})}\BibitemShut {NoStop}%
\bibitem [{\citenamefont {Rygula}\ \emph {et~al.}(1993)\citenamefont {Rygula},
  \citenamefont {Kemmler-Sack}, \citenamefont {Nissel},\ and\ \citenamefont
  {Hübener}}]{Rygula1993}%
  \BibitemOpen
  \bibfield  {author} {\bibinfo {author} {\bibfnamefont {M.}~\bibnamefont
  {Rygula}}, \bibinfo {author} {\bibfnamefont {S.}~\bibnamefont
  {Kemmler-Sack}}, \bibinfo {author} {\bibfnamefont {T.}~\bibnamefont
  {Nissel}},\ and\ \bibinfo {author} {\bibfnamefont {R.~P.}\ \bibnamefont
  {Hübener}},\ }\href {https://doi.org/10.1002/andp.19935050803} {\bibfield
  {journal} {\bibinfo  {journal} {Ann. Physik}\ }\textbf {\bibinfo {volume}
  {505}},\ \bibinfo {pages} {685} (\bibinfo {year} {1993})}\BibitemShut
  {NoStop}%
\bibitem [{\citenamefont {Moshopoulou}(1999)}]{Moshopoulou1999}%
  \BibitemOpen
  \bibfield  {author} {\bibinfo {author} {\bibfnamefont {E.~G.}\ \bibnamefont
  {Moshopoulou}},\ }\href {https://doi.org/10.1111/j.1151-2916.1999.tb02245.x}
  {\bibfield  {journal} {\bibinfo  {journal} {J. Am. Ceram. Soc.}\ }\textbf
  {\bibinfo {volume} {82}},\ \bibinfo {pages} {3317} (\bibinfo {year}
  {1999})}\BibitemShut {NoStop}%
\bibitem [{\citenamefont {Moshopoulou}\ \emph {et~al.}(1993)\citenamefont
  {Moshopoulou}, \citenamefont {Bordet}, \citenamefont {Capponi}, \citenamefont
  {Chaillout}, \citenamefont {Souletie},\ and\ \citenamefont
  {Sulpice}}]{Moshopoulou1993}%
  \BibitemOpen
  \bibfield  {author} {\bibinfo {author} {\bibfnamefont {E.}~\bibnamefont
  {Moshopoulou}}, \bibinfo {author} {\bibfnamefont {P.}~\bibnamefont {Bordet}},
  \bibinfo {author} {\bibfnamefont {J.}~\bibnamefont {Capponi}}, \bibinfo
  {author} {\bibfnamefont {C.}~\bibnamefont {Chaillout}}, \bibinfo {author}
  {\bibfnamefont {B.}~\bibnamefont {Souletie}},\ and\ \bibinfo {author}
  {\bibfnamefont {A.}~\bibnamefont {Sulpice}},\ }\href
  {https://doi.org/10.1016/0925-8388(93)90692-G} {\bibfield  {journal}
  {\bibinfo  {journal} {J. Alloy. Compd.}\ }\textbf {\bibinfo {volume} {195}},\
  \bibinfo {pages} {81 } (\bibinfo {year} {1993})}\BibitemShut {NoStop}%
\bibitem [{\citenamefont {Moshopoulou}\ \emph {et~al.}(1994)\citenamefont
  {Moshopoulou}, \citenamefont {Bordet}, \citenamefont {Sulpice},\ and\
  \citenamefont {Capponi}}]{Moshopoulou1994}%
  \BibitemOpen
  \bibfield  {author} {\bibinfo {author} {\bibfnamefont {E.}~\bibnamefont
  {Moshopoulou}}, \bibinfo {author} {\bibfnamefont {P.}~\bibnamefont {Bordet}},
  \bibinfo {author} {\bibfnamefont {A.}~\bibnamefont {Sulpice}},\ and\ \bibinfo
  {author} {\bibfnamefont {J.}~\bibnamefont {Capponi}},\ }\href
  {https://doi.org/https://doi.org/10.1016/0921-4534(94)91598-9} {\bibfield
  {journal} {\bibinfo  {journal} {Physica C}\ }\textbf {\bibinfo {volume}
  {235-240}},\ \bibinfo {pages} {747 } (\bibinfo {year} {1994})}\BibitemShut
  {NoStop}%
\bibitem [{\citenamefont {Maruyama}\ \emph {et~al.}(2015)\citenamefont
  {Maruyama}, \citenamefont {Shin}, \citenamefont {Zhang}, \citenamefont
  {Suchoski}, \citenamefont {Yasui}, \citenamefont {Jin}, \citenamefont
  {Greene},\ and\ \citenamefont {Takeuchi}}]{Maruyama2015}%
  \BibitemOpen
  \bibfield  {author} {\bibinfo {author} {\bibfnamefont {S.}~\bibnamefont
  {Maruyama}}, \bibinfo {author} {\bibfnamefont {J.}~\bibnamefont {Shin}},
  \bibinfo {author} {\bibfnamefont {X.}~\bibnamefont {Zhang}}, \bibinfo
  {author} {\bibfnamefont {R.}~\bibnamefont {Suchoski}}, \bibinfo {author}
  {\bibfnamefont {S.}~\bibnamefont {Yasui}}, \bibinfo {author} {\bibfnamefont
  {K.}~\bibnamefont {Jin}}, \bibinfo {author} {\bibfnamefont {R.~L.}\
  \bibnamefont {Greene}},\ and\ \bibinfo {author} {\bibfnamefont
  {I.}~\bibnamefont {Takeuchi}},\ }\href {https://doi.org/10.1063/1.4932551}
  {\bibfield  {journal} {\bibinfo  {journal} {Appl. Phys. Lett.}\ }\textbf
  {\bibinfo {volume} {107}},\ \bibinfo {pages} {142602} (\bibinfo {year}
  {2015})}\BibitemShut {NoStop}%
\bibitem [{\citenamefont {Jia}\ \emph {et~al.}(2018)\citenamefont {Jia},
  \citenamefont {He}, \citenamefont {Hu}, \citenamefont {Yang}, \citenamefont
  {Yang}, \citenamefont {Yu}, \citenamefont {Zhang}, \citenamefont {Shi},
  \citenamefont {Lin}, \citenamefont {Yuan}, \citenamefont {Zhu}, \citenamefont
  {Gu}, \citenamefont {Li},\ and\ \citenamefont {Jin}}]{Jia2018}%
  \BibitemOpen
  \bibfield  {author} {\bibinfo {author} {\bibfnamefont {Y.}~\bibnamefont
  {Jia}}, \bibinfo {author} {\bibfnamefont {G.}~\bibnamefont {He}}, \bibinfo
  {author} {\bibfnamefont {W.}~\bibnamefont {Hu}}, \bibinfo {author}
  {\bibfnamefont {H.}~\bibnamefont {Yang}}, \bibinfo {author} {\bibfnamefont
  {Z.}~\bibnamefont {Yang}}, \bibinfo {author} {\bibfnamefont {H.}~\bibnamefont
  {Yu}}, \bibinfo {author} {\bibfnamefont {Q.}~\bibnamefont {Zhang}}, \bibinfo
  {author} {\bibfnamefont {J.}~\bibnamefont {Shi}}, \bibinfo {author}
  {\bibfnamefont {Z.}~\bibnamefont {Lin}}, \bibinfo {author} {\bibfnamefont
  {J.}~\bibnamefont {Yuan}}, \bibinfo {author} {\bibfnamefont {B.}~\bibnamefont
  {Zhu}}, \bibinfo {author} {\bibfnamefont {L.}~\bibnamefont {Gu}}, \bibinfo
  {author} {\bibfnamefont {H.}~\bibnamefont {Li}},\ and\ \bibinfo {author}
  {\bibfnamefont {K.}~\bibnamefont {Jin}},\ }\href
  {https://doi.org/10.1038/s41598-018-22393-8} {\bibfield  {journal} {\bibinfo
  {journal} {Sci. Rep.}\ }\textbf {\bibinfo {volume} {8}},\ \bibinfo {pages}
  {3995} (\bibinfo {year} {2018})}\BibitemShut {NoStop}%
\bibitem [{\citenamefont {Yoshimatsu}\ \emph {et~al.}(2015)\citenamefont
  {Yoshimatsu}, \citenamefont {Niwa}, \citenamefont {Mashiko}, \citenamefont
  {Oshima},\ and\ \citenamefont {Ohtomo}}]{Yoshimatsu2015}%
  \BibitemOpen
  \bibfield  {author} {\bibinfo {author} {\bibfnamefont {K.}~\bibnamefont
  {Yoshimatsu}}, \bibinfo {author} {\bibfnamefont {M.}~\bibnamefont {Niwa}},
  \bibinfo {author} {\bibfnamefont {H.}~\bibnamefont {Mashiko}}, \bibinfo
  {author} {\bibfnamefont {T.}~\bibnamefont {Oshima}},\ and\ \bibinfo {author}
  {\bibfnamefont {A.}~\bibnamefont {Ohtomo}},\ }\href
  {https://doi.org/10.1038/Srep16325} {\bibfield  {journal} {\bibinfo
  {journal} {Sci. Rep.}\ }\textbf {\bibinfo {volume} {5}},\ \bibinfo {pages}
  {16325} (\bibinfo {year} {2015})}\BibitemShut {NoStop}%
\bibitem [{sup()}]{supplemental}%
  \BibitemOpen
  \href@noop {} {}\bibinfo {note} {See Supplemental Material at [URL will be
  inserted by publisher] for the calculation method for the quantity of doped
  charge, transport results of sample S5, the computational details for the
  electronic structure, and the discussion of interplay between
  superconductivity and orbital ordering, which includes Refs.
  \cite{Blochl1994,Kresse1996,Akimoto1992,Schmidt2004,Dudarev1998,Leoni2008,Hu2020,Piatti2018,Zhou2020}.}\BibitemShut
  {Stop}%
\bibitem [{\citenamefont {Lu}\ \emph {et~al.}(2017)\citenamefont {Lu},
  \citenamefont {Zhang}, \citenamefont {Zhang}, \citenamefont {Qiao},
  \citenamefont {He}, \citenamefont {Li}, \citenamefont {Wang}, \citenamefont
  {Guo}, \citenamefont {Zhang}, \citenamefont {Duan}, \citenamefont {Li},
  \citenamefont {Wang}, \citenamefont {Yang}, \citenamefont {Yan},
  \citenamefont {Arenholz}, \citenamefont {Zhou}, \citenamefont {Yang},
  \citenamefont {Gu}, \citenamefont {Nan}, \citenamefont {Wu}, \citenamefont
  {Tokura},\ and\ \citenamefont {Yu}}]{Lu2017}%
  \BibitemOpen
  \bibfield  {author} {\bibinfo {author} {\bibfnamefont {N.~P.}\ \bibnamefont
  {Lu}}, \bibinfo {author} {\bibfnamefont {P.~F.}\ \bibnamefont {Zhang}},
  \bibinfo {author} {\bibfnamefont {Q.~H.}\ \bibnamefont {Zhang}}, \bibinfo
  {author} {\bibfnamefont {R.~M.}\ \bibnamefont {Qiao}}, \bibinfo {author}
  {\bibfnamefont {Q.}~\bibnamefont {He}}, \bibinfo {author} {\bibfnamefont
  {H.~B.}\ \bibnamefont {Li}}, \bibinfo {author} {\bibfnamefont {Y.~J.}\
  \bibnamefont {Wang}}, \bibinfo {author} {\bibfnamefont {J.~W.}\ \bibnamefont
  {Guo}}, \bibinfo {author} {\bibfnamefont {D.}~\bibnamefont {Zhang}}, \bibinfo
  {author} {\bibfnamefont {Z.}~\bibnamefont {Duan}}, \bibinfo {author}
  {\bibfnamefont {Z.~L.}\ \bibnamefont {Li}}, \bibinfo {author} {\bibfnamefont
  {M.}~\bibnamefont {Wang}}, \bibinfo {author} {\bibfnamefont {S.~Z.}\
  \bibnamefont {Yang}}, \bibinfo {author} {\bibfnamefont {M.~Z.}\ \bibnamefont
  {Yan}}, \bibinfo {author} {\bibfnamefont {E.}~\bibnamefont {Arenholz}},
  \bibinfo {author} {\bibfnamefont {S.~Y.}\ \bibnamefont {Zhou}}, \bibinfo
  {author} {\bibfnamefont {W.~L.}\ \bibnamefont {Yang}}, \bibinfo {author}
  {\bibfnamefont {L.}~\bibnamefont {Gu}}, \bibinfo {author} {\bibfnamefont
  {C.~W.}\ \bibnamefont {Nan}}, \bibinfo {author} {\bibfnamefont
  {J.}~\bibnamefont {Wu}}, \bibinfo {author} {\bibfnamefont {Y.}~\bibnamefont
  {Tokura}},\ and\ \bibinfo {author} {\bibfnamefont {P.}~\bibnamefont {Yu}},\
  }\href {https://doi.org/10.1038/nature22389} {\bibfield  {journal} {\bibinfo
  {journal} {Nature (London)}\ }\textbf {\bibinfo {volume} {546}},\ \bibinfo
  {pages} {124} (\bibinfo {year} {2017})}\BibitemShut {NoStop}%
\bibitem [{\citenamefont {Perez-Mu{\~n}oz}\ \emph {et~al.}(2017)\citenamefont
  {Perez-Mu{\~n}oz}, \citenamefont {Schio}, \citenamefont {Poloni},
  \citenamefont {Fernandez-Martinez}, \citenamefont {Rivera-Calzada},
  \citenamefont {Cezar}, \citenamefont {Salas-Colera}, \citenamefont {Castro},
  \citenamefont {Kinney}, \citenamefont {Leon}, \citenamefont {Santamaria},
  \citenamefont {Garcia-Barriocanal},\ and\ \citenamefont
  {Goldman}}]{Perez2017}%
  \BibitemOpen
  \bibfield  {author} {\bibinfo {author} {\bibfnamefont {A.~M.}\ \bibnamefont
  {Perez-Mu{\~n}oz}}, \bibinfo {author} {\bibfnamefont {P.}~\bibnamefont
  {Schio}}, \bibinfo {author} {\bibfnamefont {R.}~\bibnamefont {Poloni}},
  \bibinfo {author} {\bibfnamefont {A.}~\bibnamefont {Fernandez-Martinez}},
  \bibinfo {author} {\bibfnamefont {A.}~\bibnamefont {Rivera-Calzada}},
  \bibinfo {author} {\bibfnamefont {J.~C.}\ \bibnamefont {Cezar}}, \bibinfo
  {author} {\bibfnamefont {E.}~\bibnamefont {Salas-Colera}}, \bibinfo {author}
  {\bibfnamefont {G.~R.}\ \bibnamefont {Castro}}, \bibinfo {author}
  {\bibfnamefont {J.}~\bibnamefont {Kinney}}, \bibinfo {author} {\bibfnamefont
  {C.}~\bibnamefont {Leon}}, \bibinfo {author} {\bibfnamefont {J.}~\bibnamefont
  {Santamaria}}, \bibinfo {author} {\bibfnamefont {J.}~\bibnamefont
  {Garcia-Barriocanal}},\ and\ \bibinfo {author} {\bibfnamefont {A.~M.}\
  \bibnamefont {Goldman}},\ }\href {https://doi.org/10.1073/pnas.1613006114}
  {\bibfield  {journal} {\bibinfo  {journal} {Proc. Natl. Acad. Sci. USA}\
  }\textbf {\bibinfo {volume} {114}},\ \bibinfo {pages} {215} (\bibinfo {year}
  {2017})}\BibitemShut {NoStop}%
\bibitem [{\citenamefont {Zhang}\ \emph {et~al.}(2017)\citenamefont {Zhang},
  \citenamefont {Zeng}, \citenamefont {Yin}, \citenamefont {Asmara},
  \citenamefont {Yang}, \citenamefont {Han}, \citenamefont {Cao}, \citenamefont
  {Zhou}, \citenamefont {Wan}, \citenamefont {Tang}, \citenamefont {Rusydi},
  \citenamefont {Ariando},\ and\ \citenamefont {Venkatesan}}]{Zhang2017}%
  \BibitemOpen
  \bibfield  {author} {\bibinfo {author} {\bibfnamefont {L.}~\bibnamefont
  {Zhang}}, \bibinfo {author} {\bibfnamefont {S.}~\bibnamefont {Zeng}},
  \bibinfo {author} {\bibfnamefont {X.}~\bibnamefont {Yin}}, \bibinfo {author}
  {\bibfnamefont {T.~C.}\ \bibnamefont {Asmara}}, \bibinfo {author}
  {\bibfnamefont {P.}~\bibnamefont {Yang}}, \bibinfo {author} {\bibfnamefont
  {K.}~\bibnamefont {Han}}, \bibinfo {author} {\bibfnamefont {Y.}~\bibnamefont
  {Cao}}, \bibinfo {author} {\bibfnamefont {W.}~\bibnamefont {Zhou}}, \bibinfo
  {author} {\bibfnamefont {D.}~\bibnamefont {Wan}}, \bibinfo {author}
  {\bibfnamefont {C.~S.}\ \bibnamefont {Tang}}, \bibinfo {author}
  {\bibfnamefont {A.}~\bibnamefont {Rusydi}}, \bibinfo {author} {\bibnamefont
  {Ariando}},\ and\ \bibinfo {author} {\bibfnamefont {T.}~\bibnamefont
  {Venkatesan}},\ }\href {https://doi.org/10.1021/acsnano.7b03978} {\bibfield
  {journal} {\bibinfo  {journal} {ACS Nano}\ }\textbf {\bibinfo {volume}
  {11}},\ \bibinfo {pages} {9950} (\bibinfo {year} {2017})}\BibitemShut
  {NoStop}%
\bibitem [{\citenamefont {Lei}\ \emph {et~al.}(2017)\citenamefont {Lei},
  \citenamefont {Wang}, \citenamefont {Shang}, \citenamefont {Meng},
  \citenamefont {Ma}, \citenamefont {Luo}, \citenamefont {Wu}, \citenamefont
  {Sun}, \citenamefont {Wang}, \citenamefont {Jiang}, \citenamefont {Mao},
  \citenamefont {Liu}, \citenamefont {Yu}, \citenamefont {Zhang},\ and\
  \citenamefont {Chen}}]{Lei2017}%
  \BibitemOpen
  \bibfield  {author} {\bibinfo {author} {\bibfnamefont {B.}~\bibnamefont
  {Lei}}, \bibinfo {author} {\bibfnamefont {N.~Z.}\ \bibnamefont {Wang}},
  \bibinfo {author} {\bibfnamefont {C.}~\bibnamefont {Shang}}, \bibinfo
  {author} {\bibfnamefont {F.~B.}\ \bibnamefont {Meng}}, \bibinfo {author}
  {\bibfnamefont {L.~K.}\ \bibnamefont {Ma}}, \bibinfo {author} {\bibfnamefont
  {X.~G.}\ \bibnamefont {Luo}}, \bibinfo {author} {\bibfnamefont
  {T.}~\bibnamefont {Wu}}, \bibinfo {author} {\bibfnamefont {Z.}~\bibnamefont
  {Sun}}, \bibinfo {author} {\bibfnamefont {Y.}~\bibnamefont {Wang}}, \bibinfo
  {author} {\bibfnamefont {Z.}~\bibnamefont {Jiang}}, \bibinfo {author}
  {\bibfnamefont {B.~H.}\ \bibnamefont {Mao}}, \bibinfo {author} {\bibfnamefont
  {Z.}~\bibnamefont {Liu}}, \bibinfo {author} {\bibfnamefont {Y.~J.}\
  \bibnamefont {Yu}}, \bibinfo {author} {\bibfnamefont {Y.~B.}\ \bibnamefont
  {Zhang}},\ and\ \bibinfo {author} {\bibfnamefont {X.~H.}\ \bibnamefont
  {Chen}},\ }\href {https://doi.org/10.1103/PhysRevB.95.020503} {\bibfield
  {journal} {\bibinfo  {journal} {Phys. Rev. B}\ }\textbf {\bibinfo {volume}
  {95}},\ \bibinfo {pages} {020503(R)} (\bibinfo {year} {2017})}\BibitemShut
  {NoStop}%
\bibitem [{\citenamefont {Ying}\ \emph {et~al.}(2018)\citenamefont {Ying},
  \citenamefont {Wang}, \citenamefont {Wu}, \citenamefont {Zhao}, \citenamefont
  {Zhang}, \citenamefont {Song}, \citenamefont {Li}, \citenamefont {Lei},
  \citenamefont {Li}, \citenamefont {Yu}, \citenamefont {Cheng}, \citenamefont
  {An}, \citenamefont {Zhang}, \citenamefont {Jia}, \citenamefont {Yang},
  \citenamefont {Chen},\ and\ \citenamefont {Li}}]{Ying2018}%
  \BibitemOpen
  \bibfield  {author} {\bibinfo {author} {\bibfnamefont {T.~P.}\ \bibnamefont
  {Ying}}, \bibinfo {author} {\bibfnamefont {M.~X.}\ \bibnamefont {Wang}},
  \bibinfo {author} {\bibfnamefont {X.~X.}\ \bibnamefont {Wu}}, \bibinfo
  {author} {\bibfnamefont {Z.~Y.}\ \bibnamefont {Zhao}}, \bibinfo {author}
  {\bibfnamefont {Z.~Z.}\ \bibnamefont {Zhang}}, \bibinfo {author}
  {\bibfnamefont {B.~Q.}\ \bibnamefont {Song}}, \bibinfo {author}
  {\bibfnamefont {Y.~C.}\ \bibnamefont {Li}}, \bibinfo {author} {\bibfnamefont
  {B.}~\bibnamefont {Lei}}, \bibinfo {author} {\bibfnamefont {Q.}~\bibnamefont
  {Li}}, \bibinfo {author} {\bibfnamefont {Y.}~\bibnamefont {Yu}}, \bibinfo
  {author} {\bibfnamefont {E.~J.}\ \bibnamefont {Cheng}}, \bibinfo {author}
  {\bibfnamefont {Z.~H.}\ \bibnamefont {An}}, \bibinfo {author} {\bibfnamefont
  {Y.}~\bibnamefont {Zhang}}, \bibinfo {author} {\bibfnamefont {X.~Y.}\
  \bibnamefont {Jia}}, \bibinfo {author} {\bibfnamefont {W.}~\bibnamefont
  {Yang}}, \bibinfo {author} {\bibfnamefont {X.~H.}\ \bibnamefont {Chen}},\
  and\ \bibinfo {author} {\bibfnamefont {S.~Y.}\ \bibnamefont {Li}},\ }\href
  {https://doi.org/10.1103/PhysRevLett.121.207003} {\bibfield  {journal}
  {\bibinfo  {journal} {Phys. Rev. Lett.}\ }\textbf {\bibinfo {volume} {121}},\
  \bibinfo {pages} {207003} (\bibinfo {year} {2018})}\BibitemShut {NoStop}%
\bibitem [{\citenamefont {Qin}\ \emph {et~al.}(2020)\citenamefont {Qin},
  \citenamefont {Zhang}, \citenamefont {Lin}, \citenamefont {Feng},
  \citenamefont {Wei}, \citenamefont {Alvarez}, \citenamefont {Dong},
  \citenamefont {Silhanek}, \citenamefont {Zhu}, \citenamefont {Yuan},
  \citenamefont {Qin},\ and\ \citenamefont {Jin}}]{Qin2020}%
  \BibitemOpen
  \bibfield  {author} {\bibinfo {author} {\bibfnamefont {M.~Y.}\ \bibnamefont
  {Qin}}, \bibinfo {author} {\bibfnamefont {R.~Z.}\ \bibnamefont {Zhang}},
  \bibinfo {author} {\bibfnamefont {Z.~F.}\ \bibnamefont {Lin}}, \bibinfo
  {author} {\bibfnamefont {Z.~P.}\ \bibnamefont {Feng}}, \bibinfo {author}
  {\bibfnamefont {X.~J.}\ \bibnamefont {Wei}}, \bibinfo {author} {\bibfnamefont
  {S.~B.}\ \bibnamefont {Alvarez}}, \bibinfo {author} {\bibfnamefont
  {C.}~\bibnamefont {Dong}}, \bibinfo {author} {\bibfnamefont {A.~V.}\
  \bibnamefont {Silhanek}}, \bibinfo {author} {\bibfnamefont {B.~Y.}\
  \bibnamefont {Zhu}}, \bibinfo {author} {\bibfnamefont {J.}~\bibnamefont
  {Yuan}}, \bibinfo {author} {\bibfnamefont {Q.}~\bibnamefont {Qin}},\ and\
  \bibinfo {author} {\bibfnamefont {K.}~\bibnamefont {Jin}},\ }\href
  {https://doi.org/10.1007/s10948-019-05300-8} {\bibfield  {journal} {\bibinfo
  {journal} {J. Supercond. Nov. Magn.}\ }\textbf {\bibinfo {volume} {33}},\
  \bibinfo {pages} {159} (\bibinfo {year} {2020})}\BibitemShut {NoStop}%
\bibitem [{\citenamefont {Deschanvres}\ \emph {et~al.}(1971)\citenamefont
  {Deschanvres}, \citenamefont {Raveau},\ and\ \citenamefont
  {Sekkal}}]{DESCHANVRES1971699}%
  \BibitemOpen
  \bibfield  {author} {\bibinfo {author} {\bibfnamefont {A.}~\bibnamefont
  {Deschanvres}}, \bibinfo {author} {\bibfnamefont {B.}~\bibnamefont
  {Raveau}},\ and\ \bibinfo {author} {\bibfnamefont {Z.}~\bibnamefont
  {Sekkal}},\ }\href {https://doi.org/10.1016/0025-5408(71)90103-6} {\bibfield
  {journal} {\bibinfo  {journal} {Mater. Res. Bull.}\ }\textbf {\bibinfo
  {volume} {6}},\ \bibinfo {pages} {699 } (\bibinfo {year} {1971})}\BibitemShut
  {NoStop}%
\bibitem [{\citenamefont {Saarela}\ and\ \citenamefont
  {Kusmartsev}(2017)}]{Saarela2017}%
  \BibitemOpen
  \bibfield  {author} {\bibinfo {author} {\bibfnamefont {M.}~\bibnamefont
  {Saarela}}\ and\ \bibinfo {author} {\bibfnamefont {F.}~\bibnamefont
  {Kusmartsev}},\ }\href {https://doi.org/10.1016/j.physc.2016.08.007}
  {\bibfield  {journal} {\bibinfo  {journal} {Physica C}\ }\textbf {\bibinfo
  {volume} {533}},\ \bibinfo {pages} {9 } (\bibinfo {year} {2017})}\BibitemShut
  {NoStop}%
\bibitem [{\citenamefont {Schmidt}\ \emph {et~al.}(2004)\citenamefont
  {Schmidt}, \citenamefont {Ratcliff}, \citenamefont {Radaelli}, \citenamefont
  {Refson}, \citenamefont {Harrison},\ and\ \citenamefont
  {Cheong}}]{Schmidt2004}%
  \BibitemOpen
  \bibfield  {author} {\bibinfo {author} {\bibfnamefont {M.}~\bibnamefont
  {Schmidt}}, \bibinfo {author} {\bibfnamefont {W.}~\bibnamefont {Ratcliff}},
  \bibinfo {author} {\bibfnamefont {P.~G.}\ \bibnamefont {Radaelli}}, \bibinfo
  {author} {\bibfnamefont {K.}~\bibnamefont {Refson}}, \bibinfo {author}
  {\bibfnamefont {N.~M.}\ \bibnamefont {Harrison}},\ and\ \bibinfo {author}
  {\bibfnamefont {S.~W.}\ \bibnamefont {Cheong}},\ }\href
  {https://doi.org/10.1103/PhysRevLett.92.056402} {\bibfield  {journal}
  {\bibinfo  {journal} {Phys. Rev. Lett.}\ }\textbf {\bibinfo {volume} {92}},\
  \bibinfo {pages} {056402} (\bibinfo {year} {2004})}\BibitemShut {NoStop}%
\bibitem [{\citenamefont {Hu}\ \emph {et~al.}(2020)\citenamefont {Hu},
  \citenamefont {Feng}, \citenamefont {Gong}, \citenamefont {He}, \citenamefont
  {Li}, \citenamefont {Qin}, \citenamefont {Shi}, \citenamefont {Li},
  \citenamefont {Zhang}, \citenamefont {Yuan}, \citenamefont {Zhu},
  \citenamefont {Liu}, \citenamefont {Xiang}, \citenamefont {Gu}, \citenamefont
  {Zhou}, \citenamefont {Dong}, \citenamefont {Zhao},\ and\ \citenamefont
  {Jin}}]{Hu2020}%
  \BibitemOpen
  \bibfield  {author} {\bibinfo {author} {\bibfnamefont {W.}~\bibnamefont
  {Hu}}, \bibinfo {author} {\bibfnamefont {Z.}~\bibnamefont {Feng}}, \bibinfo
  {author} {\bibfnamefont {B.-C.}\ \bibnamefont {Gong}}, \bibinfo {author}
  {\bibfnamefont {G.}~\bibnamefont {He}}, \bibinfo {author} {\bibfnamefont
  {D.}~\bibnamefont {Li}}, \bibinfo {author} {\bibfnamefont {M.}~\bibnamefont
  {Qin}}, \bibinfo {author} {\bibfnamefont {Y.}~\bibnamefont {Shi}}, \bibinfo
  {author} {\bibfnamefont {Q.}~\bibnamefont {Li}}, \bibinfo {author}
  {\bibfnamefont {Q.}~\bibnamefont {Zhang}}, \bibinfo {author} {\bibfnamefont
  {J.}~\bibnamefont {Yuan}}, \bibinfo {author} {\bibfnamefont {B.}~\bibnamefont
  {Zhu}}, \bibinfo {author} {\bibfnamefont {K.}~\bibnamefont {Liu}}, \bibinfo
  {author} {\bibfnamefont {T.}~\bibnamefont {Xiang}}, \bibinfo {author}
  {\bibfnamefont {L.}~\bibnamefont {Gu}}, \bibinfo {author} {\bibfnamefont
  {F.}~\bibnamefont {Zhou}}, \bibinfo {author} {\bibfnamefont {X.}~\bibnamefont
  {Dong}}, \bibinfo {author} {\bibfnamefont {Z.}~\bibnamefont {Zhao}},\ and\
  \bibinfo {author} {\bibfnamefont {K.}~\bibnamefont {Jin}},\ }\href
  {https://doi.org/10.1103/PhysRevB.101.220510} {\bibfield  {journal} {\bibinfo
   {journal} {Phys. Rev. B}\ }\textbf {\bibinfo {volume} {101}},\ \bibinfo
  {pages} {220510(R)} (\bibinfo {year} {2020})}\BibitemShut {NoStop}%
\bibitem [{\citenamefont {Khomskii}\ and\ \citenamefont
  {Mizokawa}(2005)}]{Khomskii2005}%
  \BibitemOpen
  \bibfield  {author} {\bibinfo {author} {\bibfnamefont {D.~I.}\ \bibnamefont
  {Khomskii}}\ and\ \bibinfo {author} {\bibfnamefont {T.}~\bibnamefont
  {Mizokawa}},\ }\href {https://doi.org/10.1103/PhysRevLett.94.156402}
  {\bibfield  {journal} {\bibinfo  {journal} {Phys. Rev. Lett.}\ }\textbf
  {\bibinfo {volume} {94}},\ \bibinfo {pages} {156402} (\bibinfo {year}
  {2005})}\BibitemShut {NoStop}%
\bibitem [{\citenamefont {Akimoto}\ \emph {et~al.}(1992)\citenamefont
  {Akimoto}, \citenamefont {Gotoh}, \citenamefont {Kawaguchi},\ and\
  \citenamefont {Oosawa}}]{Akimoto1992}%
  \BibitemOpen
  \bibfield  {author} {\bibinfo {author} {\bibfnamefont {J.}~\bibnamefont
  {Akimoto}}, \bibinfo {author} {\bibfnamefont {Y.}~\bibnamefont {Gotoh}},
  \bibinfo {author} {\bibfnamefont {K.}~\bibnamefont {Kawaguchi}},\ and\
  \bibinfo {author} {\bibfnamefont {Y.}~\bibnamefont {Oosawa}},\ }\href
  {https://doi.org/10.1016/S0022-4596(05)80280-4} {\bibfield  {journal}
  {\bibinfo  {journal} {J. Solid State Chem.}\ }\textbf {\bibinfo {volume}
  {96}},\ \bibinfo {pages} {446 } (\bibinfo {year} {1992})}\BibitemShut
  {NoStop}%
\bibitem [{\citenamefont {Zhu}\ \emph {et~al.}(2007)\citenamefont {Zhu},
  \citenamefont {Tang}, \citenamefont {Zhao}, \citenamefont {Wang},
  \citenamefont {Li}, \citenamefont {Yin}, \citenamefont {Yu}, \citenamefont
  {Tang}, \citenamefont {Xiong}, \citenamefont {Shi},\ and\ \citenamefont
  {Ruan}}]{ZHU2007578}%
  \BibitemOpen
  \bibfield  {author} {\bibinfo {author} {\bibfnamefont {B.-P.}\ \bibnamefont
  {Zhu}}, \bibinfo {author} {\bibfnamefont {Z.}~\bibnamefont {Tang}}, \bibinfo
  {author} {\bibfnamefont {L.-H.}\ \bibnamefont {Zhao}}, \bibinfo {author}
  {\bibfnamefont {L.-L.}\ \bibnamefont {Wang}}, \bibinfo {author}
  {\bibfnamefont {C.-Z.}\ \bibnamefont {Li}}, \bibinfo {author} {\bibfnamefont
  {D.}~\bibnamefont {Yin}}, \bibinfo {author} {\bibfnamefont {Z.-X.}\
  \bibnamefont {Yu}}, \bibinfo {author} {\bibfnamefont {W.-F.}\ \bibnamefont
  {Tang}}, \bibinfo {author} {\bibfnamefont {R.}~\bibnamefont {Xiong}},
  \bibinfo {author} {\bibfnamefont {J.}~\bibnamefont {Shi}},\ and\ \bibinfo
  {author} {\bibfnamefont {X.-F.}\ \bibnamefont {Ruan}},\ }\href
  {https://doi.org/10.1016/j.matlet.2006.05.038} {\bibfield  {journal}
  {\bibinfo  {journal} {Mater. Lett.}\ }\textbf {\bibinfo {volume} {61}},\
  \bibinfo {pages} {578 } (\bibinfo {year} {2007})}\BibitemShut {NoStop}%
\bibitem [{\citenamefont {Satpathy}\ and\ \citenamefont
  {Martin}(1987)}]{Satpathy1987}%
  \BibitemOpen
  \bibfield  {author} {\bibinfo {author} {\bibfnamefont {S.}~\bibnamefont
  {Satpathy}}\ and\ \bibinfo {author} {\bibfnamefont {R.~M.}\ \bibnamefont
  {Martin}},\ }\href {https://doi.org/10.1103/PhysRevB.36.7269} {\bibfield
  {journal} {\bibinfo  {journal} {Phys. Rev. B}\ }\textbf {\bibinfo {volume}
  {36}},\ \bibinfo {pages} {7269} (\bibinfo {year} {1987})}\BibitemShut
  {NoStop}%
\bibitem [{\citenamefont {Leoni}\ \emph {et~al.}(2008)\citenamefont {Leoni},
  \citenamefont {Yaresko}, \citenamefont {Perkins}, \citenamefont {Rosner},\
  and\ \citenamefont {Craco}}]{Leoni2008}%
  \BibitemOpen
  \bibfield  {author} {\bibinfo {author} {\bibfnamefont {S.}~\bibnamefont
  {Leoni}}, \bibinfo {author} {\bibfnamefont {A.~N.}\ \bibnamefont {Yaresko}},
  \bibinfo {author} {\bibfnamefont {N.}~\bibnamefont {Perkins}}, \bibinfo
  {author} {\bibfnamefont {H.}~\bibnamefont {Rosner}},\ and\ \bibinfo {author}
  {\bibfnamefont {L.}~\bibnamefont {Craco}},\ }\href
  {https://doi.org/10.1103/PhysRevB.78.125105} {\bibfield  {journal} {\bibinfo
  {journal} {Phys. Rev. B}\ }\textbf {\bibinfo {volume} {78}},\ \bibinfo
  {pages} {125105} (\bibinfo {year} {2008})}\BibitemShut {NoStop}%
\bibitem [{\citenamefont {Zhang}\ \emph {et~al.}(2002)\citenamefont {Zhang},
  \citenamefont {Israel}, \citenamefont {Biswas}, \citenamefont {Greene},\ and\
  \citenamefont {de~Lozanne}}]{Zhang2002}%
  \BibitemOpen
  \bibfield  {author} {\bibinfo {author} {\bibfnamefont {L.}~\bibnamefont
  {Zhang}}, \bibinfo {author} {\bibfnamefont {C.}~\bibnamefont {Israel}},
  \bibinfo {author} {\bibfnamefont {A.}~\bibnamefont {Biswas}}, \bibinfo
  {author} {\bibfnamefont {R.~L.}\ \bibnamefont {Greene}},\ and\ \bibinfo
  {author} {\bibfnamefont {A.}~\bibnamefont {de~Lozanne}},\ }\href
  {https://doi.org/10.1126/science.1077346} {\bibfield  {journal} {\bibinfo
  {journal} {Science}\ }\textbf {\bibinfo {volume} {298}},\ \bibinfo {pages}
  {805} (\bibinfo {year} {2002})}\BibitemShut {NoStop}%
\bibitem [{\citenamefont {Sipos}\ \emph {et~al.}(2008)\citenamefont {Sipos},
  \citenamefont {Kusmartseva}, \citenamefont {Akrap}, \citenamefont {Berger},
  \citenamefont {Forro},\ and\ \citenamefont {Tutis}}]{Sipos2008}%
  \BibitemOpen
  \bibfield  {author} {\bibinfo {author} {\bibfnamefont {B.}~\bibnamefont
  {Sipos}}, \bibinfo {author} {\bibfnamefont {A.~F.}\ \bibnamefont
  {Kusmartseva}}, \bibinfo {author} {\bibfnamefont {A.}~\bibnamefont {Akrap}},
  \bibinfo {author} {\bibfnamefont {H.}~\bibnamefont {Berger}}, \bibinfo
  {author} {\bibfnamefont {L.}~\bibnamefont {Forro}},\ and\ \bibinfo {author}
  {\bibfnamefont {E.}~\bibnamefont {Tutis}},\ }\href
  {https://doi.org/10.1038/nmat2318} {\bibfield  {journal} {\bibinfo  {journal}
  {Nat. Mater.}\ }\textbf {\bibinfo {volume} {7}},\ \bibinfo {pages} {960}
  (\bibinfo {year} {2008})}\BibitemShut {NoStop}%
\bibitem [{\citenamefont {Mattoni}\ \emph {et~al.}(2016)\citenamefont
  {Mattoni}, \citenamefont {Zubko}, \citenamefont {Maccherozzi}, \citenamefont
  {van~der Torren}, \citenamefont {Boltje}, \citenamefont {Hadjimichael},
  \citenamefont {Manca}, \citenamefont {Catalano}, \citenamefont {Gibert},
  \citenamefont {Liu}, \citenamefont {Aarts}, \citenamefont {Triscone},
  \citenamefont {Dhesi},\ and\ \citenamefont {Caviglia}}]{Mattoni2016}%
  \BibitemOpen
  \bibfield  {author} {\bibinfo {author} {\bibfnamefont {G.}~\bibnamefont
  {Mattoni}}, \bibinfo {author} {\bibfnamefont {P.}~\bibnamefont {Zubko}},
  \bibinfo {author} {\bibfnamefont {F.}~\bibnamefont {Maccherozzi}}, \bibinfo
  {author} {\bibfnamefont {A.~J.~H.}\ \bibnamefont {van~der Torren}}, \bibinfo
  {author} {\bibfnamefont {D.~B.}\ \bibnamefont {Boltje}}, \bibinfo {author}
  {\bibfnamefont {M.}~\bibnamefont {Hadjimichael}}, \bibinfo {author}
  {\bibfnamefont {N.}~\bibnamefont {Manca}}, \bibinfo {author} {\bibfnamefont
  {S.}~\bibnamefont {Catalano}}, \bibinfo {author} {\bibfnamefont
  {M.}~\bibnamefont {Gibert}}, \bibinfo {author} {\bibfnamefont
  {Y.}~\bibnamefont {Liu}}, \bibinfo {author} {\bibfnamefont {J.}~\bibnamefont
  {Aarts}}, \bibinfo {author} {\bibfnamefont {J.~M.}\ \bibnamefont {Triscone}},
  \bibinfo {author} {\bibfnamefont {S.~S.}\ \bibnamefont {Dhesi}},\ and\
  \bibinfo {author} {\bibfnamefont {A.~D.}\ \bibnamefont {Caviglia}},\ }\href
  {https://doi.org/10.1038/ncomms13141} {\bibfield  {journal} {\bibinfo
  {journal} {Nat. Commun.}\ }\textbf {\bibinfo {volume} {7}},\ \bibinfo {pages}
  {13141} (\bibinfo {year} {2016})}\BibitemShut {NoStop}%
\bibitem [{\citenamefont {Haverkort}\ \emph {et~al.}(2005)\citenamefont
  {Haverkort}, \citenamefont {Hu}, \citenamefont {Tanaka}, \citenamefont
  {Reichelt}, \citenamefont {Streltsov}, \citenamefont {Korotin}, \citenamefont
  {Anisimov}, \citenamefont {Hsieh}, \citenamefont {Lin}, \citenamefont {Chen},
  \citenamefont {Khomskii},\ and\ \citenamefont {Tjeng}}]{Haverkort2005}%
  \BibitemOpen
  \bibfield  {author} {\bibinfo {author} {\bibfnamefont {M.~W.}\ \bibnamefont
  {Haverkort}}, \bibinfo {author} {\bibfnamefont {Z.}~\bibnamefont {Hu}},
  \bibinfo {author} {\bibfnamefont {A.}~\bibnamefont {Tanaka}}, \bibinfo
  {author} {\bibfnamefont {W.}~\bibnamefont {Reichelt}}, \bibinfo {author}
  {\bibfnamefont {S.~V.}\ \bibnamefont {Streltsov}}, \bibinfo {author}
  {\bibfnamefont {M.~A.}\ \bibnamefont {Korotin}}, \bibinfo {author}
  {\bibfnamefont {V.~I.}\ \bibnamefont {Anisimov}}, \bibinfo {author}
  {\bibfnamefont {H.~H.}\ \bibnamefont {Hsieh}}, \bibinfo {author}
  {\bibfnamefont {H.-J.}\ \bibnamefont {Lin}}, \bibinfo {author} {\bibfnamefont
  {C.~T.}\ \bibnamefont {Chen}}, \bibinfo {author} {\bibfnamefont {D.~I.}\
  \bibnamefont {Khomskii}},\ and\ \bibinfo {author} {\bibfnamefont {L.~H.}\
  \bibnamefont {Tjeng}},\ }\href
  {https://doi.org/10.1103/PhysRevLett.95.196404} {\bibfield  {journal}
  {\bibinfo  {journal} {Phys. Rev. Lett.}\ }\textbf {\bibinfo {volume} {95}},\
  \bibinfo {pages} {196404} (\bibinfo {year} {2005})}\BibitemShut {NoStop}%
\bibitem [{\citenamefont {Zhou}\ \emph {et~al.}(2006)\citenamefont {Zhou},
  \citenamefont {Li}, \citenamefont {Luo}, \citenamefont {Ma}, \citenamefont
  {Wu}, \citenamefont {Zhu}, \citenamefont {Tang}, \citenamefont {Shi},\ and\
  \citenamefont {Wang}}]{Zhou2006}%
  \BibitemOpen
  \bibfield  {author} {\bibinfo {author} {\bibfnamefont {J.}~\bibnamefont
  {Zhou}}, \bibinfo {author} {\bibfnamefont {G.}~\bibnamefont {Li}}, \bibinfo
  {author} {\bibfnamefont {J.~L.}\ \bibnamefont {Luo}}, \bibinfo {author}
  {\bibfnamefont {Y.~C.}\ \bibnamefont {Ma}}, \bibinfo {author} {\bibfnamefont
  {D.}~\bibnamefont {Wu}}, \bibinfo {author} {\bibfnamefont {B.~P.}\
  \bibnamefont {Zhu}}, \bibinfo {author} {\bibfnamefont {Z.}~\bibnamefont
  {Tang}}, \bibinfo {author} {\bibfnamefont {J.}~\bibnamefont {Shi}},\ and\
  \bibinfo {author} {\bibfnamefont {N.~L.}\ \bibnamefont {Wang}},\ }\href
  {https://doi.org/10.1103/PhysRevB.74.245102} {\bibfield  {journal} {\bibinfo
  {journal} {Phys. Rev. B}\ }\textbf {\bibinfo {volume} {74}},\ \bibinfo
  {pages} {245102} (\bibinfo {year} {2006})}\BibitemShut {NoStop}%
\bibitem [{\citenamefont {Liang}\ \emph {et~al.}(2020)\citenamefont {Liang},
  \citenamefont {Lee}, \citenamefont {Yu}, \citenamefont {Zhang}, \citenamefont
  {Liang}, \citenamefont {Zavalij}, \citenamefont {Chen}, \citenamefont
  {James}, \citenamefont {Bendersky}, \citenamefont {Davydov}, \citenamefont
  {Zhang},\ and\ \citenamefont {Takeuchi}}]{Liang2020}%
  \BibitemOpen
  \bibfield  {author} {\bibinfo {author} {\bibfnamefont {Y.~G.}\ \bibnamefont
  {Liang}}, \bibinfo {author} {\bibfnamefont {S.}~\bibnamefont {Lee}}, \bibinfo
  {author} {\bibfnamefont {H.~S.}\ \bibnamefont {Yu}}, \bibinfo {author}
  {\bibfnamefont {H.~R.}\ \bibnamefont {Zhang}}, \bibinfo {author}
  {\bibfnamefont {Y.~J.}\ \bibnamefont {Liang}}, \bibinfo {author}
  {\bibfnamefont {P.~Y.}\ \bibnamefont {Zavalij}}, \bibinfo {author}
  {\bibfnamefont {X.}~\bibnamefont {Chen}}, \bibinfo {author} {\bibfnamefont
  {R.~D.}\ \bibnamefont {James}}, \bibinfo {author} {\bibfnamefont {L.~A.}\
  \bibnamefont {Bendersky}}, \bibinfo {author} {\bibfnamefont {A.~V.}\
  \bibnamefont {Davydov}}, \bibinfo {author} {\bibfnamefont {X.~H.}\
  \bibnamefont {Zhang}},\ and\ \bibinfo {author} {\bibfnamefont
  {I.}~\bibnamefont {Takeuchi}},\ }\href
  {https://doi.org/10.1038/s41467-020-17351-w} {\bibfield  {journal} {\bibinfo
  {journal} {Nat. Commun.}\ }\textbf {\bibinfo {volume} {11}},\ \bibinfo
  {pages} {3539} (\bibinfo {year} {2020})}\BibitemShut {NoStop}%
\bibitem [{\citenamefont {Catalano}\ \emph {et~al.}(2015)\citenamefont
  {Catalano}, \citenamefont {Gibert}, \citenamefont {Bisogni}, \citenamefont
  {He}, \citenamefont {Sutarto}, \citenamefont {Viret}, \citenamefont {Zubko},
  \citenamefont {Scherwitzl}, \citenamefont {Sawatzky}, \citenamefont
  {Schmitt},\ and\ \citenamefont {Triscone}}]{Catalano2015}%
  \BibitemOpen
  \bibfield  {author} {\bibinfo {author} {\bibfnamefont {S.}~\bibnamefont
  {Catalano}}, \bibinfo {author} {\bibfnamefont {M.}~\bibnamefont {Gibert}},
  \bibinfo {author} {\bibfnamefont {V.}~\bibnamefont {Bisogni}}, \bibinfo
  {author} {\bibfnamefont {F.}~\bibnamefont {He}}, \bibinfo {author}
  {\bibfnamefont {R.}~\bibnamefont {Sutarto}}, \bibinfo {author} {\bibfnamefont
  {M.}~\bibnamefont {Viret}}, \bibinfo {author} {\bibfnamefont
  {P.}~\bibnamefont {Zubko}}, \bibinfo {author} {\bibfnamefont
  {R.}~\bibnamefont {Scherwitzl}}, \bibinfo {author} {\bibfnamefont {G.~A.}\
  \bibnamefont {Sawatzky}}, \bibinfo {author} {\bibfnamefont {T.}~\bibnamefont
  {Schmitt}},\ and\ \bibinfo {author} {\bibfnamefont {J.~M.}\ \bibnamefont
  {Triscone}},\ }\href {https://doi.org/10.1063/1.4919803} {\bibfield
  {journal} {\bibinfo  {journal} {APL Mater.}\ }\textbf {\bibinfo {volume}
  {3}},\ \bibinfo {pages} {062506} (\bibinfo {year} {2015})}\BibitemShut
  {NoStop}%
\bibitem [{\citenamefont {Leininger}\ \emph {et~al.}(2011)\citenamefont
  {Leininger}, \citenamefont {Chernyshov}, \citenamefont {Bosak}, \citenamefont
  {Berger},\ and\ \citenamefont {Inosov}}]{Leininger2011}%
  \BibitemOpen
  \bibfield  {author} {\bibinfo {author} {\bibfnamefont {P.}~\bibnamefont
  {Leininger}}, \bibinfo {author} {\bibfnamefont {D.}~\bibnamefont
  {Chernyshov}}, \bibinfo {author} {\bibfnamefont {A.}~\bibnamefont {Bosak}},
  \bibinfo {author} {\bibfnamefont {H.}~\bibnamefont {Berger}},\ and\ \bibinfo
  {author} {\bibfnamefont {D.~S.}\ \bibnamefont {Inosov}},\ }\href
  {https://doi.org/10.1103/PhysRevB.83.233101} {\bibfield  {journal} {\bibinfo
  {journal} {Phys. Rev. B}\ }\textbf {\bibinfo {volume} {83}},\ \bibinfo
  {pages} {233101} (\bibinfo {year} {2011})}\BibitemShut {NoStop}%
\bibitem [{\citenamefont {Yu}\ \emph {et~al.}(2015)\citenamefont {Yu},
  \citenamefont {Yang}, \citenamefont {Lu}, \citenamefont {Yan}, \citenamefont
  {Cho}, \citenamefont {Ma}, \citenamefont {Niu}, \citenamefont {Kim},
  \citenamefont {Son}, \citenamefont {Feng}, \citenamefont {Li}, \citenamefont
  {Cheong}, \citenamefont {Chen},\ and\ \citenamefont {Zhang}}]{Yu2015}%
  \BibitemOpen
  \bibfield  {author} {\bibinfo {author} {\bibfnamefont {Y.~J.}\ \bibnamefont
  {Yu}}, \bibinfo {author} {\bibfnamefont {F.~Y.}\ \bibnamefont {Yang}},
  \bibinfo {author} {\bibfnamefont {X.~F.}\ \bibnamefont {Lu}}, \bibinfo
  {author} {\bibfnamefont {Y.~J.}\ \bibnamefont {Yan}}, \bibinfo {author}
  {\bibfnamefont {Y.~H.}\ \bibnamefont {Cho}}, \bibinfo {author} {\bibfnamefont
  {L.~G.}\ \bibnamefont {Ma}}, \bibinfo {author} {\bibfnamefont {X.~H.}\
  \bibnamefont {Niu}}, \bibinfo {author} {\bibfnamefont {S.}~\bibnamefont
  {Kim}}, \bibinfo {author} {\bibfnamefont {Y.~W.}\ \bibnamefont {Son}},
  \bibinfo {author} {\bibfnamefont {D.~L.}\ \bibnamefont {Feng}}, \bibinfo
  {author} {\bibfnamefont {S.~Y.}\ \bibnamefont {Li}}, \bibinfo {author}
  {\bibfnamefont {S.~W.}\ \bibnamefont {Cheong}}, \bibinfo {author}
  {\bibfnamefont {X.~H.}\ \bibnamefont {Chen}},\ and\ \bibinfo {author}
  {\bibfnamefont {Y.~B.}\ \bibnamefont {Zhang}},\ }\href
  {https://doi.org/10.1038/nnano.2014.323} {\bibfield  {journal} {\bibinfo
  {journal} {Nat. Nanotechnol.}\ }\textbf {\bibinfo {volume} {10}},\ \bibinfo
  {pages} {270} (\bibinfo {year} {2015})}\BibitemShut {NoStop}%
\bibitem [{\citenamefont {Isobe}\ and\ \citenamefont {Ueda}(2002)}]{Isobe2002}%
  \BibitemOpen
  \bibfield  {author} {\bibinfo {author} {\bibfnamefont {M.}~\bibnamefont
  {Isobe}}\ and\ \bibinfo {author} {\bibfnamefont {Y.}~\bibnamefont {Ueda}},\
  }\href {https://doi.org/10.1143/JPSJ.71.1848} {\bibfield  {journal} {\bibinfo
   {journal} {J. Phys. Soc. Jpn.}\ }\textbf {\bibinfo {volume} {71}},\ \bibinfo
  {pages} {1848} (\bibinfo {year} {2002})}\BibitemShut {NoStop}%
\bibitem [{\citenamefont {Piatti}\ \emph {et~al.}(2018)\citenamefont {Piatti},
  \citenamefont {Chen}, \citenamefont {Tortello}, \citenamefont {Ye},\ and\
  \citenamefont {Gonnelli}}]{Piatti2018}%
  \BibitemOpen
  \bibfield  {author} {\bibinfo {author} {\bibfnamefont {E.}~\bibnamefont
  {Piatti}}, \bibinfo {author} {\bibfnamefont {Q.~H.}\ \bibnamefont {Chen}},
  \bibinfo {author} {\bibfnamefont {M.}~\bibnamefont {Tortello}}, \bibinfo
  {author} {\bibfnamefont {J.~T.}\ \bibnamefont {Ye}},\ and\ \bibinfo {author}
  {\bibfnamefont {R.~S.}\ \bibnamefont {Gonnelli}},\ }\href
  {https://doi.org/10.1016/j.apsusc.2018.05.232} {\bibfield  {journal}
  {\bibinfo  {journal} {Appl. Surf. Sci.}\ }\textbf {\bibinfo {volume} {461}},\
  \bibinfo {pages} {269} (\bibinfo {year} {2018})}\BibitemShut {NoStop}%
\bibitem [{\citenamefont {Zhou}\ \emph {et~al.}(2020)\citenamefont {Zhou},
  \citenamefont {Deng}, \citenamefont {Guo},\ and\ \citenamefont
  {Guo}}]{Zhou2020}%
  \BibitemOpen
  \bibfield  {author} {\bibinfo {author} {\bibfnamefont {K.~Y.}\ \bibnamefont
  {Zhou}}, \bibinfo {author} {\bibfnamefont {J.}~\bibnamefont {Deng}}, \bibinfo
  {author} {\bibfnamefont {L.~W.}\ \bibnamefont {Guo}},\ and\ \bibinfo {author}
  {\bibfnamefont {J.~G.}\ \bibnamefont {Guo}},\ }\href
  {https://doi.org/10.1088/0256-307X/37/9/097402} {\bibfield  {journal}
  {\bibinfo  {journal} {Chin. Phys. Lett.}\ }\textbf {\bibinfo {volume} {37}},\
  \bibinfo {pages} {097402} (\bibinfo {year} {2020})}\BibitemShut {NoStop}%
\bibitem [{\citenamefont {Kreisel}\ \emph {et~al.}(2017)\citenamefont
  {Kreisel}, \citenamefont {Andersen}, \citenamefont {Sprau}, \citenamefont
  {Kostin}, \citenamefont {Davis},\ and\ \citenamefont
  {Hirschfeld}}]{Kreisel2017}%
  \BibitemOpen
  \bibfield  {author} {\bibinfo {author} {\bibfnamefont {A.}~\bibnamefont
  {Kreisel}}, \bibinfo {author} {\bibfnamefont {B.~M.}\ \bibnamefont
  {Andersen}}, \bibinfo {author} {\bibfnamefont {P.~O.}\ \bibnamefont {Sprau}},
  \bibinfo {author} {\bibfnamefont {A.}~\bibnamefont {Kostin}}, \bibinfo
  {author} {\bibfnamefont {J.~C.~S.}\ \bibnamefont {Davis}},\ and\ \bibinfo
  {author} {\bibfnamefont {P.~J.}\ \bibnamefont {Hirschfeld}},\ }\href
  {https://doi.org/10.1103/PhysRevB.95.174504} {\bibfield  {journal} {\bibinfo
  {journal} {Phys. Rev. B}\ }\textbf {\bibinfo {volume} {95}},\ \bibinfo
  {pages} {174504} (\bibinfo {year} {2017})}\BibitemShut {NoStop}%
\bibitem [{\citenamefont {Sprau}\ \emph {et~al.}(2017)\citenamefont {Sprau},
  \citenamefont {Kostin}, \citenamefont {Kreisel}, \citenamefont {Bohmer},
  \citenamefont {Taufour}, \citenamefont {Canfield}, \citenamefont {Mukherjee},
  \citenamefont {Hirschfeld}, \citenamefont {Andersen},\ and\ \citenamefont
  {Davis}}]{Sprau2017}%
  \BibitemOpen
  \bibfield  {author} {\bibinfo {author} {\bibfnamefont {P.~O.}\ \bibnamefont
  {Sprau}}, \bibinfo {author} {\bibfnamefont {A.}~\bibnamefont {Kostin}},
  \bibinfo {author} {\bibfnamefont {A.}~\bibnamefont {Kreisel}}, \bibinfo
  {author} {\bibfnamefont {A.~E.}\ \bibnamefont {Bohmer}}, \bibinfo {author}
  {\bibfnamefont {V.}~\bibnamefont {Taufour}}, \bibinfo {author} {\bibfnamefont
  {P.~C.}\ \bibnamefont {Canfield}}, \bibinfo {author} {\bibfnamefont
  {S.}~\bibnamefont {Mukherjee}}, \bibinfo {author} {\bibfnamefont {P.~J.}\
  \bibnamefont {Hirschfeld}}, \bibinfo {author} {\bibfnamefont {B.~M.}\
  \bibnamefont {Andersen}},\ and\ \bibinfo {author} {\bibfnamefont {J.~C.~S.}\
  \bibnamefont {Davis}},\ }\href {https://doi.org/10.1126/science.aal1575}
  {\bibfield  {journal} {\bibinfo  {journal} {Science}\ }\textbf {\bibinfo
  {volume} {357}},\ \bibinfo {pages} {75} (\bibinfo {year} {2017})}\BibitemShut
  {NoStop}%
\bibitem [{\citenamefont {Bl\"ochl}(1994)}]{Blochl1994}%
  \BibitemOpen
  \bibfield  {author} {\bibinfo {author} {\bibfnamefont {P.~E.}\ \bibnamefont
  {Bl\"ochl}},\ }\href {https://doi.org/10.1103/PhysRevB.50.17953} {\bibfield
  {journal} {\bibinfo  {journal} {Phys. Rev. B}\ }\textbf {\bibinfo {volume}
  {50}},\ \bibinfo {pages} {17953} (\bibinfo {year} {1994})}\BibitemShut
  {NoStop}%
\bibitem [{\citenamefont {Kresse}\ and\ \citenamefont
  {Furthm\"uller}(1996)}]{Kresse1996}%
  \BibitemOpen
  \bibfield  {author} {\bibinfo {author} {\bibfnamefont {G.}~\bibnamefont
  {Kresse}}\ and\ \bibinfo {author} {\bibfnamefont {J.}~\bibnamefont
  {Furthm\"uller}},\ }\href {https://doi.org/10.1016/0927-0256(96)00008-0}
  {\bibfield  {journal} {\bibinfo  {journal} {Comput. Mater. Sci.}\ }\textbf
  {\bibinfo {volume} {6}},\ \bibinfo {pages} {15 } (\bibinfo {year}
  {1996})}\BibitemShut {NoStop}%
\bibitem [{\citenamefont {Dudarev}\ \emph {et~al.}(1998)\citenamefont
  {Dudarev}, \citenamefont {Botton}, \citenamefont {Savrasov}, \citenamefont
  {Humphreys},\ and\ \citenamefont {Sutton}}]{Dudarev1998}%
  \BibitemOpen
  \bibfield  {author} {\bibinfo {author} {\bibfnamefont {S.~L.}\ \bibnamefont
  {Dudarev}}, \bibinfo {author} {\bibfnamefont {G.~A.}\ \bibnamefont {Botton}},
  \bibinfo {author} {\bibfnamefont {S.~Y.}\ \bibnamefont {Savrasov}}, \bibinfo
  {author} {\bibfnamefont {C.~J.}\ \bibnamefont {Humphreys}},\ and\ \bibinfo
  {author} {\bibfnamefont {A.~P.}\ \bibnamefont {Sutton}},\ }\href
  {https://doi.org/10.1103/PhysRevB.57.1505} {\bibfield  {journal} {\bibinfo
  {journal} {Phys. Rev. B}\ }\textbf {\bibinfo {volume} {57}},\ \bibinfo
  {pages} {1505} (\bibinfo {year} {1998})}\BibitemShut {NoStop}%
\end{thebibliography}%

\end{document}